\documentclass[twocolumn,pra,aps,showpacs,groupedaddress]{revtex4}
\usepackage{graphics}
\usepackage{amsmath}
\newcommand{\la}{\langle}
\newcommand{\ra}{\rangle}
\newcommand{\kv}{{k_\mathrm{ v}}}
\newcommand{\re}{\mathrm{ Re}}
\begin{document}
\title{Ghost imaging using homodyne detection}

\author{M. Bache} 
\email[Corresponding author: ]{morten.bache@uninsubria.it}
\affiliation{INFM, Dipartimento di Scienze, Universit{\`a} dell'Insubria,
  Via Valleggio 11, 22100 Como, Italy}
\author{E. Brambilla}
\affiliation{INFM, Dipartimento di Scienze, Universit{\`a} dell'Insubria,
  Via Valleggio 11, 22100 Como, Italy}
\author{A. Gatti}
\affiliation{INFM, Dipartimento di Scienze, Universit{\`a} dell'Insubria,
  Via Valleggio 11, 22100 Como, Italy}
\author{L.A. Lugiato} 
\affiliation{INFM, Dipartimento di Scienze, Universit{\`a} dell'Insubria,
  Via Valleggio 11, 22100 Como, Italy}

 \date{\today}
\begin{abstract}
  We present a theoretical study of ghost imaging based on correlated
  beams arising from parametric down-conversion, and which uses
  balanced homodyne detection to measure both the signal and idler
  fields. We analytically show that the signal-idler correlations
  contain the full amplitude and phase information about an object
  located in the signal path, both in the near-field and the far-field
  case. To this end we discuss how to optimize the optical setups in
  the two imaging paths, including the crucial point regarding how to
  engineer the phase of the idler local oscillator as to observe the
  desired orthogonal quadrature components of the image. We point out
  an inherent link between the far-field bandwidth and the near-field
  resolution of the reproduced image, determined by the bandwidth of
  the source of the correlated beams. However, we show how to
  circumvent this limitation by using a spatial averaging technique
  which dramatically improves the imaging bandwidth of the far-field
  correlations as well as speeds up the convergence rate. The results
  are backed up by numerical simulations taking into account the
  finite size and duration of the pump pulse.
\end{abstract}

\maketitle
%

\section{Introduction}
\label{sec:Introduction}

Ghost imaging relies on the spatial correlation between two beams
created by, e.g., parametric down conversion (PDC)
\cite{klyshko:1988,belinskii:1994,strekalov:1995,pittman:1995,ribiero:1994,ribeiro:1999,saleh:2000,abouraddy:2001,abouraddy:2002,abouraddy:2003,gatti:2003,gatti:2004,gatti:2004b,bennink:2002a,bennink:2004}.
Each of the correlated beams are sent through a distinct imaging
system called the test arm and the reference arm.  In the test arm an
object is placed and the image of the object is then recreated from
the spatial correlation function between the test and reference arm.
A basic requisite of the ghost-imaging schemes is that by solely
adjusting the reference arm setup and varying the reference
point-like detector position it should be possible to retrieve
spatial information about the object, such as the object image (near field)
and the object diffraction pattern (far field).

Traditionally the studies devoted to ghost imaging have considered PDC
in the low-gain regime, where the conversion rate of the pump photons
to a pair of entangled signal-idler photons is low enough for the
detector to resolve them one at a time.  This leads to the so-called
two-photon imaging schemes first investigated by Klyshko
\cite{klyshko:1988,belinskii:1994}.  The recreation of the object is
then based on coincidence counts between the test and reference arm,
and the working principles have been demonstrated experimentally
\cite{strekalov:1995,pittman:1995,ribiero:1994}. Our group has focused
on generalizing the governing PDC theory to the macroscopic, high-gain
regime where the number of photons per mode is large
\cite{gatti:1999b,brambilla:2001,navez:2002,gatti:2003c,brambilla:2004}.
Moreover, we generalized the theory behind the two-photon imaging
schemes
\cite{klyshko:1988,belinskii:1994,ribeiro:1999,saleh:2000,abouraddy:2001,abouraddy:2002}
to the high-gain regime \cite{gatti:2003,gatti:2004,gatti:2004b} and
also investigated the quantum properties of the signal-idler
correlations in that case
\cite{brambilla:2001,navez:2002,gatti:2003c,brambilla:2004}.  While
coincidence detection is used for low gain, in the high-gain regime
the object information is extracted from measuring pixel-by-pixel the
signal-idler intensities and from this forming the correlations.

Here we consider a setup where each of the PDC beams are measured with
balanced homodyne detection by overlapping them with local oscillator
(LO) fields. A similar scheme was studied previously by our group in
the context of entanglement of the signal-idler beams from PDC
\cite{navez:2002}, where it was used to show analytically that the
entanglement is complete since it encompasses both the amplitudes and
the phases of signal and idler.  The initial motivation for using a
homodyne scheme for quantum imaging came from the need to circumvent
the problems related to information visibility in the macroscopic
regime. Specifically, when intensity detection is performed a
homogeneous background term is present in the measured correlation
function \cite{gatti:2003,gatti:2004,brambilla:2004}. This term,
which can be rather large, does not contain any information about the
object and lowers the image visibility. Instead, by using homodyne
detection the signal-idler correlation becomes second order instead of
fourth order, and hence this background term is absent.  Another
advantage of homodyne detection is that arbitrary quadrature
components of the test and reference beams can be measured, which
means that the homodyne detection scheme allows for both amplitude and
phase measurements of the object.  We will show detailed analytical
calculations which demonstrate that this is indeed possible, both for
the object image (near field) as well as its diffraction pattern (far
field). It is possible to reconstruct even a pure phase object when a
bucket detector is used in the test arm (in contrast to when intensity
measurements are done, as it was shown for the coincidence counting
case \cite{bennink:2002a}).

We will present a technique that implements an average over the test
detector position, and this turns out to strongly improve the imaging
bandwidth of the system with respect to when a fixed test detector
position is used, where the imaging bandwidth is limited to the
far-field bandwidth of PDC. Additionally the method substantially
speeds up the convergence rate of the far-field correlations. This
implies that even complex diffraction patterns with high-frequency
Fourier components can be reconstructed in a low number of pump
pulses.

The suggested homodyne detection scheme is fairly complicated to
implement experimentally, since it involves carrying out two
independent homodyne measurements of fields that are spatially and
temporally multimode (with pulse durations on the order of ps). A dual
homodyne measurement has been done in the case of spatially
single-mode light \cite{ou:1992,kuzmich:2000}, but even a single
homodyne measurement of spatially multimode light has not previously
been investigated in details to our knowledge, not even in the
continuous wave regime where the field is temporally single-mode. 
Homodyne experiments of spatially single-mode but temporally multimode
light have pointed out the importance of a proper overlap between the
LO and the field \cite{slusher:1987,yurke:1987}, and alternative ways
of producing the LO has been implemented \cite{kim:1994} leading to a
better overlap.  An improper overlap leads to a degrading in, e.g.,
the squeezing as it effectively corresponds to a drop in the quantum
efficiency \cite{grosshans:2001}.  In our case these problems are
presumably less severe since the quantum efficiency does not play an
important role.  In fact, we will present numerical simulations that
confirm that homodyne measurement of spatio-temporal multimode light
can be done successfully, with the strongest technical restriction
concerning the engineering of the phases of the local oscillators. It
is worth to mention that the homodyne measurement protocol (no
background term in the correlations) in combination with the spatial
averaging technique (increased spatial bandwidth) could open for the
possibility of using an OPO below threshold for imaging. This would
simplify a possible experimental implementation.

In Sec.~\ref{sec:model} the model for PDC is presented, and the the
optical setup of the imaging system is described in
Sec.~\ref{sec:system-setup}. In Sec.~\ref{sec:Point-like-detector} we
discuss the case where point-like detectors are used in the test arm,
while in Sec.~\ref{sec:Bucket-detector-at} the bucket detector case is
discussed, and show analytical results in the stationary and plane-wave
pump approximation concerning the retrieval of the image information
of both the far and the near field, as well as numerical results that
include the Gaussian profile of the pump. Besides confirming the
analytical results these serve as examples for discussion and
demonstration of the imaging performances of the system. In
Sec.~\ref{sec:Conclusion} we draw the conclusions. The appendices
discuss a quadratic expansion of the PDC gain phase
(App.~\ref{sec:Append-appr-form}), and how to treat the temporal part
of the analytical correlations (App.~\ref{sec:sign-idler-corr}).

\section{The model}
\label{sec:model}

The starting point is a general model describing the three-wave
quantum interaction of PDC inside the nonlinear crystal
\cite{gatti:2003c,brambilla:2004}, which includes the effects of
finite size and duration of the pump and the effects of
spatio-temporal walk-off and group-velocity dispersion.

We consider a uniaxial $\chi^{(2)}$ nonlinear crystal cut for type II
phase matching. The injected beam at the frequency $\omega_0$ (pump beam)
is sent into the crystal in the $z$ direction. Inside the crystal the
pump photons may then down-convert to sets of photons at the
frequencies $\omega_1$ (signal photon, ordinarily polarized) and
$\omega_2$ (idler photon, extraordinarily polarized). The classical
equations governing PDC are in the paraxial and slowly varying envelope
approximation 
\begin{subequations}
  \label{eq:classicPDC}
\begin{eqnarray}
  {\mathcal L}_j A_j&=&\sigma A_0 A^*_l e^{-i\Delta_0z}, \quad j,l=1,2,
  \quad j\neq l \label{eq:signal-idlerPDC}\\
  {\mathcal L}_0 A_0&=&-\sigma A_1 A_2 e^{i\Delta_0z} .
\end{eqnarray}
\end{subequations}
${\mathcal L}_j$ are operators describing linear propagation in the
medium, including walk-off as well as 2nd order dispersion effects in
the spatial and temporal domains 
\begin{equation}
  \label{eq:Lj}
  {\mathcal L}_j = \frac{\partial}{\partial z}
  +k_j'\frac{\partial}{\partial t} 
  +\frac{ik_j''}{2}\frac{\partial^2}{\partial t^2}  
  -\rho_j\frac{\partial}{\partial x} 
  -\frac{i}{2k_j}\nabla_\perp^2.
\end{equation}
$k_j=n_j\omega_j/c$ is the wave number of field $j$ inside the
crystal, and the primes denote derivatives with respect to $\omega$
taken at the carrier frequency $\omega_j$: the two terms describe
temporal walk-off ($k_j'=\partial k_j/\partial
\omega|_{\omega=\omega_j}$) and group velocity dispersion
($k_j''=\partial^2 k_j/\partial \omega^2|_{\omega=\omega_j}$).
Dealing for simplicity with uniaxial crystals, we have assumed that
the walk-off direction of the extraordinary waves is along the
$x$-axis. The walk-off angle $\rho_j=\partial k/\partial
q_x|_{\vec{q}=0}$ determines the energy propagation direction of wave
$j$ with respect to the pump axis $z$. The effect of diffraction is
described in the framework of the paraxial approximation through the
Laplacian $\nabla_\perp^2=\partial^2/\partial x^2+\partial^2/\partial
y^2$ with $\vec{x}=(x,y)$ spanning the transverse plane.

The right hand sides of Eq.~(\ref{eq:classicPDC}) give the nonlinear
interaction in the crystal, the strength of the nonlinearity being
governed by $\sigma$ that is proportional to the effective second
order susceptibility $\chi^{(2)}_\mathrm{eff}$. $\Delta_0\equiv
k_1+k_2-k_0$ denotes the collinear phase-mismatch between the three
waves along the $z$-axis.

These classical equations can be converted into a set of operator
equations describing the evolution of the quantum mechanical
operators. This conversion is formally done by substituting
$A_j(\vec{x},z,t)\rightarrow a_j(\vec{x},z,t)$, $j=1,2$. 
$a_j(\vec{x},z,t)$ are boson operators obeying the following
commutator relations at a given $z$
\begin{eqnarray}
\label{eq:commut}
[a_i(z,\vec{x},t),a_j^{\dag}(z,\vec{x}',t')]
&=&\delta_{ij}\delta(\vec{x}-\vec{x}')\delta(t-t'),
\end{eqnarray}
$j=1,2$, while all other combinations commute. With this choice the
expectation value of the intensity $a^\dagger a$ gives the number of
photons per time per area. The pump $A_0$ is treated classically, and
furthermore we adopt the parametric approximation where the pump is
taken as undepleted. It is more convenient for the subsequent analysis
to present the equations in Fourier space according to the following
transformation
\begin{equation}
a_j(z,\vec{q},\Omega)=
     \int\frac{d\vec{x}}{2\pi}\int\frac{dt}{\sqrt{2\pi}}
      a_j(z,\vec{x},t)e^{-i\vec{q}\cdot\vec{x}+i\Omega t},
\end{equation}
$j=1,2$. The signal equation~(\ref{eq:signal-idlerPDC}) is then
\begin{eqnarray}
\label{eq:waveq}
\tilde{\mathcal L}_1(z,\vec{q},\Omega) a_1(z,\vec{q},\Omega)=\sigma
e^{-i\Delta_0 z} 
    \int\frac{d\vec{q}'}{2\pi} \int\frac{d\Omega'}{\sqrt{2\pi}}
\nonumber\\\times
         A_0(z,\vec{q}-\vec{q}',\Omega-\Omega')
    a_2^{\dag}(z,-\vec{q}',-\Omega')      ,
\end{eqnarray}
while the idler equation can be found by exchanging subscripts
$1\leftrightarrow 2$. The Fourier version of the linear propagation
operator is defined as $\tilde{\mathcal L}_j(z,\vec{q},\Omega)\equiv
\frac{\partial}{\partial z}- i\delta_j(\vec{q},\Omega)$, where the
advantage of being in the Fourier space is that the derivatives with
respect to $t$ and $\vec{x}$ become constants
\begin{equation}
\label{eq:kz}
\delta_j(\vec{q},\Omega)
=k_j'\Omega+\frac{1}{2}k_j''\Omega^2+\rho_j q_x-
\frac{1}{2k_j}|\vec{q}|^2 .
\end{equation}

The classical pump is taken as being Gaussian in both time and space
and can be expressed as
\begin{subequations}
\begin{eqnarray}
  \label{eq:pumpnear}
  A_0(z=0,\vec{x},t)&=&(2\pi)^{3/2}A_pe^{-|\vec{x}|^2/w_0^2-t^2/\tau_0^2},\\
  A_0(z=0,\vec{q},\Omega)&=&\frac{2\sqrt{2}A_p}
  {\delta q_0^2\delta\omega_0}e^{-|\vec{q}|^2/\delta q_0^2-
  \Omega^2/\delta\omega_0^2} , 
\end{eqnarray}
\end{subequations}
in the real and Fourier space, respectively. Here $w_0$ and $\tau_0$
are the pump waist and duration time, while $A_p$ is the pump
amplitude. The bandwidths in Fourier space are given by $\delta
q_0=2/w_0$ and $\delta \omega_0=2/\tau_0$.

In the stationary and plane-wave pump approximation (SPWPA) the pump is
taken as translationally invariant as well as continuous wave. Thus,
$w_0$ and $\tau_0$ tend to infinity so we have
$A_0(z,\vec{q},\Omega)\rightarrow
(2\pi)^{3/2}A_p~\delta(\vec{q})\delta(\Omega)$. Under this condition
Eqs.~(\ref{eq:waveq}) can be solved analytically. The unitary
input-output transformations relating the field operators at the
output face of the crystal $a_j^{\mathrm{out}}(\vec{q},\Omega)\equiv
a_j(z=l_c,\vec{q},\Omega)$ to those at the input face
$a_j^{\mathrm{in}}(\vec{q},\Omega)\equiv a_j(z=0,\vec{q},\Omega)$ then
take the following form
\begin{eqnarray}
a_{i}^\mathrm{out}(\vec{q},\Omega)&=&U_i(\vec{q},\Omega)a_{i}^\mathrm{
  in} (\vec{q},\Omega)
\nonumber\\
&+&V_i(\vec{q},\Omega)a_{j}^\mathrm{ in\dag}(-\vec{q},-\Omega), \quad
  i\neq j=1,2.
\label{eq:inputoutput}
\end{eqnarray}
The signal gain functions are 
\begin{subequations}
\label{eq:uv}
\begin{eqnarray}
     U_1(\vec{q},\Omega)=&
     e^{iD_{12}(\vec{q},\Omega)l_c/2}
     \Big[\cosh(\Gamma_{12}(\vec{q},\Omega)l_c)
\nonumber\\
     &+i\frac{\Delta_{12}(\vec{q},\Omega)}{2\Gamma_{12}(\vec{q},\Omega)}
     \sinh(\Gamma_{12}(\vec{q},\Omega)l_c)\Big],\\
     V_1(\vec{q},\Omega)=&
      e^{iD_{12}(\vec{q},\Omega)l_c/2}
     \frac{\sigma_p\sinh(\Gamma_{12}(\vec{q},\Omega)l_c)}
     {\Gamma_{12}(\vec{q},\Omega)}  
     .
\end{eqnarray}
\end{subequations}
For the idler similar gain functions are found by exchanging
indices $1\leftrightarrow 2$. We have in Eq.~(\ref{eq:uv}) introduced
$\sigma_p\equiv\sigma A_p$ as well as
\begin{subequations}
\label{Gamma}
\begin{eqnarray}
D_{ij}(\vec{q},\Omega)&\equiv&\delta_i(\vec{q},\Omega)-
      \delta_j(-\vec{q},-\Omega)-\Delta_0,\\
      \Gamma_{ij}(\vec{q},\Omega)
      &\equiv&\sqrt{\sigma_p^2-[\Delta_{ij}(\vec{q},\Omega)/2]^2}\;,\\
      \Delta_{ij}(\vec{q},\Omega) \label{eq:Deltaq}
      &\equiv&\Delta_0+\delta_i(\vec{q},\Omega)+\delta_j(-\vec{q},-\Omega).
\end{eqnarray}
\end{subequations}
It is important to note that the gain functions satisfy the following
unitarity conditions
\begin{subequations}
\label{eq:unitarity}
\begin{eqnarray}
     |U_j(\vec{q},\Omega)|^2-|V_j(\vec{q},\Omega)|^2=1,
     \quad (j=1,2)\label{unit1}\\  
     U_1(\vec{q},\Omega)V_2(-\vec{q},-\Omega)=
     U_2(-\vec{q},-\Omega)V_1(\vec{q},\Omega),\label{unit2}
\end{eqnarray}
\end{subequations}
which guarantee the conservation of the free-field commutation
relations~(\ref{eq:commut}) after propagation.

The bandwidths of emission of the gain functions~(\ref{eq:uv}) in the
spatial and temporal frequency domain are
\begin{subequations}
\label{eq:coherence}
\begin{eqnarray}
  \label{eq:q0}
  q_0^2&=&2[l_c(1/k_1+1/k_2)]^{-1},\\
  \label{eq:omega0}
  \Omega_0&=&(|k_1'-k_2'|l_c)^{-1},
\end{eqnarray}
\end{subequations} 
where the typical variation scales of the signal-idler fields in space
and time are
\begin{equation}
  x_\mathrm{coh}=1/q_0,\quad \tau_\mathrm{
  coh}=1/\Omega_0.
\label{eq:coherence-space-time}
\end{equation}
They can be identified with the coherence length and coherence time,
respectively, of the fields.

\section{The system setup}
\label{sec:system-setup}

\begin{figure}[t]
\begin{center}
{\scalebox{.4}{\includegraphics*{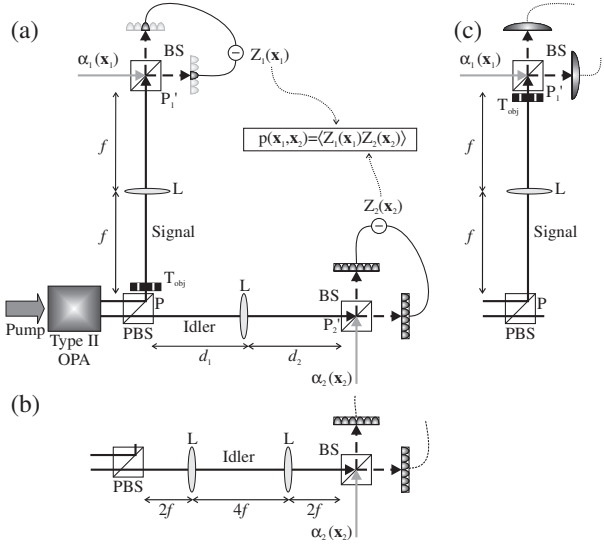}}}
\caption{The system setup.   Sketch (a) shows the general setup where
  a point-like detector (or a pixel array) is used in the test
  (signal) arm. Sketch (b) shows the case of a telescope setup in
  the reference (idler) arm.  Sketch (c) shows the test arm setup when
  bucket detectors are used for the measurements there. See the text for
  more details. L: lens of focal length $f$, PBS: polarized beam
  splitter, BS: beam splitter, $P$: crystal plane, $P_1'$: signal
  measurement plane, $P_2'$: idler measurement plane.  }
\label{fig:setup}
\end{center}
\end{figure}

Ghost imaging is characterized by its two-arm configuration, with an
unknown object placed in the test arm. The information about the
object is retrieved from the cross-correlations of the fields recorded
in the test and reference arms as a function of the reference arm pixel
position, see Fig.~\ref{fig:setup}. By simply changing the optical
setup in the reference arm, information about both the object image
(near field) and the object diffraction pattern (far field, i.e. the
Fourier transform) can be obtained. There are two main motivations for
using such an imaging configuration. Firstly, this configuration
makes it possible to do coherent imaging even if each of the two
fields are spatially incoherent (and thus recording the far-field
spatial distribution in the test arm does not give any information
about the object diffraction pattern). Secondly, the
configuration allows for a simple detection protocol in the test arm
even without measuring spatial information, and still the spatial
information can be retrieved from the correlations. Thus a simple
bucket detector setup can be used that collects all photons, or
alternatively using a single pixel detector (point-like detector). This
is advantageous when the object is located in a environment that is
hard to access making it difficult to place an array of detectors
after the object, or if the stability of the test arm is an issue making
a simple setup crucial. 

In this publication we will focus mainly on the point-like detector in
the signal arm for two reasons. First of all, in the high-gain regime
there are many photons per mode implying that a single-pixel
measurement carry enough information to perform the correlation. In
contrast, in the low-gain regime the probability of having a photon
after the object is so low that a bucket detector is sometimes favored
in order to get coincidences within a reasonable time. Secondly,
having spatial resolution in the signal arm allows to perform averages
over the pixels of the test arm detector (in this case it is necessary
to have an array of pixels in the test arm), which as we will show
vastly speeds up the convergence rate as well as improves the imaging
bandwidth. For completeness we also treat the bucket detector case in
Sec.~\ref{sec:Bucket-detector-at}.

We consider the case where the signal and idler input fields are in
the vacuum state. Thus, in the SPWPA we have the following
correlations at the crystal exit \cite{gatti:2003c,brambilla:2004}
\begin{subequations}
\label{eq:corr_crystal_exit}
\begin{eqnarray}
  \la a_{j}^\mathrm{ out\dagger}(\vec{q},\Omega)a_{k}^\mathrm{
  out}(\vec{q}',\Omega') \ra \nonumber\\= \delta_{jk} \delta
  (\vec{q}-\vec{q}')\delta(\Omega-\Omega')|V_j(\vec{q},\Omega)|^2,
  \quad j=1,2\\ 
  \la a_{1}^\mathrm{ out}(\vec{q},\Omega)a_{2}^\mathrm{
  out}(\vec{q}',\Omega') \ra \nonumber\\=
  \delta (\vec{q}+\vec{q}')\delta(\Omega+\Omega')U_1(\vec{q},\Omega)
  V_2(-\vec{q},-\Omega)  .
\label{eq:a1a2-corr}
\end{eqnarray}
\end{subequations}

The type II phase matching conditions of the crystal ensure that the
signal and idler have orthogonal polarizations and therefore at the
exit of the crystal they can be separated by a polarizing beam
splitter (PBS). We will neglect the distance between the crystal and
the PBS. The fields $c_j(\vec{x},t)$ at the measurement planes $P_j'$
are connected to the output fields at the crystal exit plane $P$ by a
Fresnel transformation formally written as
\begin{eqnarray}
  \label{eq:cj}
  c_j(\vec{x}_j,t) &=&\int d\vec{x}'h_{j}(\vec{x}_j,\vec{x}')
a_j^\mathrm{ out}(\vec{x}',t)+V_j(\vec{x}_j),
\end{eqnarray}
where $h_j(\vec{x}_j,\vec{x}')$ are kernels related to the optical
path from $P$ to $P_j'$.  $V_j(\vec{x}_j)$ accounts for losses in
the imaging system that are linearly proportional to vacuum field
operators and therefore uncorrelated to the PDC fields. 

The object is described by a transmission function $T_\mathrm{
  obj}(\vec{x})$, and is set in the signal arm. Besides the object the
signal arm is set in the so-called f-f imaging scheme, consisting of a
lens L with focal length $f$ located at a distance $f$ from both the
crystal exit and from the detection plane $P_1'$.  If we apply the
unfolded Klyshko picture \cite{klyshko:1988} it is obvious that the
required position of the object changes depending on the type of
detector used in the test arm. Namely, when a point-like detector is
used the detector acts like a point-like source which through the f-f
lens system becomes plane wave source at the crystal, and therefore
the optimal object placement is at $P$, as shown in
Fig.~\ref{fig:setup}a.  Conversely, if a bucket detector is chosen it
acts like a plane wave source at $P_1'$ and through the f-f lens
system it changes to a point-like source at $P$.  Thus, if in this
case we put the object at $P$ we will not get any information about
it, while if we put the object at $P_1'$ it is instead effectively
illuminated by the plane wave source.  Thus, in the bucket case the
object must be placed at $P_1'$, as shown in Fig.~\ref{fig:setup}c.
Consequently, the signal Fresnel kernels are
\begin{subequations}
\begin{eqnarray}
  \label{eq:h1}
h_{1}(\vec{x}_1,\vec{x}')&=&(i \lambda_1 f)^{-1}
e^{-i\frac{2\pi}{\lambda_1 f}\vec{x}_1\cdot\vec{x}'}T_\mathrm{
  obj}(\vec{x}'),\\
  \label{eq:h1-bucket-xx'}
h_{1}^\mathrm{b}(\vec{x}_1,\vec{x}')&=&(i \lambda_1 f)^{-1}
e^{-i\frac{2\pi}{\lambda_1 f}\vec{x}_1\cdot\vec{x}'}T_\mathrm{
  obj}(\vec{x}_1),
\end{eqnarray}
\end{subequations}
for the point-like and the bucket detector case, respectively, and the
superscript ``b'' denotes bucket.  $\lambda_1$ is the wavelength of
the signal field.  Note that the lens in the signal arm could actually
be removed and substituted by a propagation over a distance long
enough to be in the Frauenhofer regime. However, the phase of the
signal field could be difficult to control in that case.

We will fix the setup of the signal arm once and for all. We may then
adjust the idler arm to retrieve the desired kind of information
through the signal-idler correlations. The idler lens system also
consists of a lens L of focal length $f$, as shown in
Fig.~\ref{fig:setup}a. In one setup $d_1$ and $d_2$ are identical and
both equal to the lens focal length, $d_1=d_2=f$, i.e. an f-f setup as
also used in the signal arm. With this setup the correlations in the
point-like detector case give information about the object far field,
while in the bucket detector case they give information about the
object near field. In the other setup, instead, these distances obey
the thin lens law $1/d_1+1/d_2=1/f$. This means that the object near
field can be retrieved from the correlations in the point-like
detector case, while the object far field can be retrieved in the
bucket detector case. For simplicity we may keep $d_1=d_2=2f$ and we
denote this the 2f-2f setup.  We will also consider the so-called
telescope case, see Fig.~\ref{fig:setup}b, where the 2f-2f setup is
repeated twice adding an extra lens, but the system still obeys
the thin lens law: therefore the object information can be retrieved
as in the 2f-2f case. For these three chosen setups in the idler
plane, the idler field at $P_2'$ is then given by the Fresnel
transformation~(\ref{eq:cj}) with the kernels
\begin{subequations}
\label{eq:idler}
\begin{eqnarray}
h_{2,f}(\vec{x},\vec{x}')&=&(i\lambda_2 f)^{-1}
e^{-i\frac{2\pi}{\lambda_2 f}\vec{x}\cdot\vec{x}'}\;
,\label{eq:h2_f}\\
h_{2,2f}(\vec{x},\vec{x}')&=& e^{-i\frac{\pi}{\lambda_2
    f}|\vec{x}|^2} \delta( 
\vec{x}+\vec{x}'),\label{eq:h2_2f}\\
h_{2,T}(\vec{x},\vec{x}')&=& \delta(\vec{x}-\vec{x}'),
\label{eq:h2_T}
\end{eqnarray}
\end{subequations}
for the f-f setup, the 2f-2f setup, and the telescope setup, respectively. 
We choose to work with the telescope setup with respect to the 2f-2f
setup because of the direct transformation from the image plane $P$ to
the measurement plane $P_2'$ [due to the $\delta(\vec{x}'-\vec{x})$
type correlation], as well as the lack of the phase factor. 

The fields at $P_j'$ are measured using balanced homodyne detection
schemes, so $c_j(\vec{x}_j,t)$ is mixed with an LO
$\alpha_j(\vec{x}_j,t)$ on a 50/50 beam splitter, and after the beam
splitter the two fields are
\begin{equation}
  \label{eq:cj_pm}
  c_{j,\pm}(\vec{x},t)=\frac{c_j(\vec{x},t)\pm
  \alpha_j(\vec{x},t)}{\sqrt{2}} .
\end{equation}
The LO is treated as a classical coherent field, therefore by
measuring the field intensities at the photo-detectors and subtracting
them we obtain
\begin{eqnarray}
  Z_j(\vec{x},t)&\equiv& c_{j,+}^{\dagger}(\vec{x},t)c_{j,+}(\vec{x},t)
  - c_{j,-}^\dagger(\vec{x},t)c_{j,-}(\vec{x},t) \nonumber\\
&=&c_j(\vec{x},t)\alpha_j^*(\vec{x},t)
  + c_j^\dagger(\vec{x},t)\alpha_j(\vec{x},t)   
\nonumber
\\
&=&|\alpha_j(\vec{x},t)|
\nonumber\\ &\times&
[c_j(\vec{x},t)e^{-i\phi_j^\mathrm{
  LO}(\vec{x},t)} 
+ c_j^\dagger(\vec{x},t)e^{i\phi_j^\mathrm{
  LO}(\vec{x},t)} ].   \label{eq:quad}
\end{eqnarray}
Here we have the well known result that by properly adjusting the local
oscillator phase $\phi_j^\mathrm{LO}(\vec{x},t)$ one can measure a
particular quadrature component of the field.

We now consider the correlation between two particular quadrature
components of the signal and idler fields, $Z_1$ and $Z_2$, integrated
over a finite detection time $T_d$ 
\begin{subequations}
\begin{eqnarray}
  \label{eq:p12}
  p(\vec{x}&_1,\vec{x}_2) =\int_{T_d} dt_1 \int_{T_d} dt_2 \la
  Z_1(\vec{x}_1,t_1) 
  Z_2(\vec{x}_2,t_2) \ra 
\\
&=\int_{T_d} dt_1 \int_{T_d} dt_2
 \alpha_1^*(\vec{x}_1,t_1) 
 \alpha_2^*(\vec{x}_2,t_2)
\nonumber\\
&\times\la  c_1(\vec{x}_1,t_1) c_2(\vec{x}_2,t_2) \ra
 +\mathrm{c.c.}
  \label{eq:p12_c}
\end{eqnarray}
\end{subequations}
The last line follows from adopting the parametric approximation as
well as from the fact that the signal and idler are in the vacuum
state at the crystal entrance, since under these assumptions $\la
c_1^\dagger(\vec{x}_1,t_1) c_2(\vec{x}_2,t_2) \ra=0$. In most cases
$T_d\gg \tau_\mathrm{coh}$ so the detector is too slow to follow the
temporal dynamics of the system. This means that the measured
intensities are in fact proportional to the time-integrated field
quadratures selected by the LOs.

In the SPWPA we may evaluate the correlation~(\ref{eq:p12_c}) using
the formalism outlined in Eqs.~(\ref{eq:inputoutput})-(\ref{Gamma}).
We will do this in the degenerate case where signal and idler have
identical wavelengths $\lambda_1=\lambda_2\equiv \lambda$ and hence
also vacuum wave numbers $k_{1,\mathrm v}=k_{2,\mathrm v}\equiv
\kv=2\pi/\lambda$.  In order to evaluate the
correlation~(\ref{eq:p12_c}) we note that Eq.~(\ref{eq:cj}) provides
the Fresnel transformation from the exit of the crystal to the
detection plane. The correlations at the crystal exit are the near
field signal-idler correlation which can be found
from~(\ref{eq:a1a2-corr}) as
\begin{eqnarray}
  \label{eq:a1a2-q}
  \la a_{1}^{\mathrm{out}}(\vec{x}_1,t_1)
  a_{2}^{\mathrm{out}}(\vec{x}_2,t_2)\ra =
\nonumber\\
  \int \frac{d\Omega}{2\pi} e^{-i\Omega(t_1-t_2)}
\int \frac{d \vec{q}}{(2\pi)^2}e^{i \vec{q}(\vec{x}_1-\vec{x}_2)}
  G(\vec{q},\Omega),
\end{eqnarray}
where we have introduced the gain function
\begin{equation}
  \label{eq:G}
 G(\vec{q},\Omega)=|G(\vec{q},\Omega)|e^{i\phi_G(\vec{q},\Omega)}
\equiv U_1(\vec{q},\Omega) 
V_2(-\vec{q},-\Omega) .
\end{equation}
Appendix~\ref{sec:Append-appr-form} shows an approximate quadratic
expansion of the gain phase that will become useful in the following.

We now use (\ref{eq:cj}) and (\ref{eq:a1a2-q}) to obtain a more useful
notation for the signal-idler correlation at the detectors
\begin{subequations}
\begin{eqnarray}
  \label{eq:c1c2-u1u2}
  \la c_1(\vec{x}_1,t_1)c_2(\vec{x}_2,t_2)\ra =  \int \frac{d\Omega}{2\pi}
  e^{-i\Omega(t_1-t_2)}
\nonumber\\
\times\int d\vec{q}
h_1(\vec{x}_1,-\vec{q}) h_2(\vec{x}_2,\vec{q})G(\vec{q},\Omega)   ,
\\
    \label{eq:hxq}
h_j(\vec{x},\vec{q})\equiv \int \frac{d\vec{x}'}{2\pi} e^{-i
  \vec{q}\cdot \vec{x}'} h_j(\vec{x},\vec{x}').
\end{eqnarray}
\end{subequations}
By using Eq.~(\ref{eq:c1c2-u1u2}) we may evaluate the signal-idler
correlation~(\ref{eq:p12_c}) for $T_d\gg \tau_\mathrm{coh}$. The
details of this calculation are put in
Appendix~\ref{sec:sign-idler-corr}, and basically there are two limits
to consider. One is the case of a pulsed LO where the LO duration time
is much smaller than the coherence time, and the result is given by
Eq.~(\ref{eq:p12-pulsed}).  The other case is that of a
continuous-wave (cw) LO, where the LO duration time is much larger
than the coherence time, and the result is given by
Eq.~(\ref{eq:p12-mono-app}). As mentioned in the appendix we did not
find substantial differences between the two cases, and we therefore
decided to use the latter in the rest of this paper for the analytical
calculations. For completeness we repeat Eq.~(\ref{eq:p12-mono-app}),
stating that for a cw LO the following expression holds
\begin{eqnarray}
  \label{eq:p12-mono}
  p(\vec{x}_1,\vec{x}_2)=\frac{T_d}{2\pi}
\alpha_1^*(\vec{x}_1)\alpha_2^*(\vec{x}_2)
\nonumber\\\times
  \int d\vec{q} 
h_1(\vec{x}_1,-\vec{q}) h_2(\vec{x}_2,\vec{q})  G(\vec{q},0) 
  +\mathrm{c.c.},
\end{eqnarray}
where $\alpha_j(\vec{x})\equiv\alpha_j(\vec{x},t=0)$.

In the following two sections we treat the two choices for the
measurement technique in the signal arm, namely point-like detectors
and bucket detectors. In each section both analytical and numerical
results are presented.

\section{Point-like detector setup in test arm}
\label{sec:Point-like-detector}

When point-like detectors are used in the test arm the setup of
Fig.~\ref{fig:setup}a is used. The signal arm we keep fixed in the f-f
setup with the object at the crystal exit. The idler arm is then set
in either the f-f setup or the telescope setup. The information is
extracted by fixing the signal detector position $\vec{x}_1$ and
scanning the idler detector position $\vec{x}_2$. This gives a simple,
fixed setup of the test arm.

\subsection{Analytical results}
\label{sec:Analytical-results}

The analytical results are based on the SPWPA, so evaluating
(\ref{eq:hxq}) for the signal we use~(\ref{eq:h1}) to obtain
\begin{eqnarray}
h_1(\vec{x}_1,\vec{q})=(i \lambda f)^{-1}\tilde{T}_\mathrm{
  obj}( \vec{x}_1\kv/f+\vec{q}),
  \label{eq:Gamma1-final}
\end{eqnarray}
where $\tilde{T}_\mathrm{obj}(\vec{q})$ is the Fourier transform of
$T_\mathrm{obj}(\vec{x})$. 

\subsubsection{Retrieval of object far field}
\label{sec:Retrieval-object-far}

The object far field can be retrieved from the correlations in the
point-like detector case by properly adjusting the optical setup in
the reference arm based on the f-f scheme. We therefore
evaluate~(\ref{eq:hxq}) in the f-f case where by using~(\ref{eq:h2_f})
we obtain
\begin{equation}
  \label{eq:h2q-ff}
  h_{2,f}(\vec{x}_2,\vec{q})=2\pi(i\lambda f)^{-1}
\delta (\vec{q}+\vec {x}_2 \kv/f).
\end{equation}
By inserting~(\ref{eq:Gamma1-final}) and~(\ref{eq:h2q-ff})
in~(\ref{eq:p12-mono}) the signal-idler quadrature correlation is
\begin{subequations}
\begin{eqnarray}
  p_f(\vec{x}_1,\vec{x}_2)&=&\frac{2T_d}{(\lambda f)^2}|G(-\vec{x}_2\kv/f,0)
    \alpha_1(\vec{x}_1)\alpha_2(\vec{x}_2)| 
\nonumber\\&\times &
\re\left(\tilde T_\mathrm{obj} [(\vec{x}_1+\vec{x}_2)\kv/f]
  e^{i\Phi_f(\vec{x}_1,\vec{x}_2)}\right),
  \label{eq:p_f}
\\
\Phi_f(\vec{x}_1,\vec{x}_2)&\equiv&
\phi_G(-\vec{x}_2\kv/f,0)
\nonumber\\
&&-\phi_{2,f}^\mathrm{
  LO}(\vec{x}_2) -\phi_1^\mathrm{LO}(\vec{x}_1) +\pi.
\label{eq:phi_f}
\end{eqnarray}
\end{subequations}
This is the main result for the retrieval of the object far field. We
now discuss how to optimize the system for obtaining the far-field
image information. 

First of all we notice that we need to keep the spatial dependence of
the LO moduli as constant as possible, effectively setting a limit on
their spatial waists. However, since the effect of the LO moduli is a
simple multiplication on the correlation it is easy to do a
post-measurement correction of the correlations. This means that
matching the spatial overlaps of the LO moduli and the PDC fields
should be straightforward, while the phases of the LOs are more
crucial, as discussed below. The same comments turn out to hold for
the other setups discussed in the paper.  

Second of all, when $\vec{x}_2$ is scanned the gain modulus comes into
play. We choose a non-collinear phase-matching condition (see
Sec.~\ref{sec:Numerical-results}) in which
$|G(\vec{q}=-\vec{x}_2\kv/f,0)|$ is a plateau shaped function centered
on $\vec{q}=-\vec{q}_C$, with \cite{brambilla:2004}
\begin{equation}
  \label{eq:qC}
\vec{q}_C=\frac{1}{2}\vec{\rho}_2l_cq_0^2=\vec{\rho}_2(1/k_1+1/k_2)^{-1}.
\end{equation} 
Additionally, it is roughly constant over a finite region determined
by the spatial bandwidth $q_0$ defined by Eq.~(\ref{eq:q0}). This sets
a limit to the imaging resolution since the higher order spatial
frequency components are cut off by the gain. In
Sec.~\ref{sec:Averaging-over-shots-1} we will show how to circumvent
this limitation.

Finally, the most crucial point is to control the LO phases to observe
the desired quadrature components of the object far field. It is seen
that the real part of the object far field is obtained if we can make
$\Phi_f(\vec{x}_1,\vec{x_2})=0$, while if we can make
$\Phi_f(\vec{x}_1,\vec{x_2})=-\pi/2$ the imaginary part of the object
far field is obtained. This can be achieved by engineering
$\phi_{2,f}^\mathrm{LO}(\vec{x}_2)$ to cancel the gain phase
dependence on $\vec{x}_2$ that appears in Eq.~(\ref{eq:phi_f}). As a
good approximation we may take $\phi_{2,f}^\mathrm{LO}(\vec{x}_2)$
constant because the gain phase is a slow function over a region
determined by $q_0$, and we may therefore evaluate it anywhere inside
the gain region, e.g. at the gain center $-\vec{x}_2\kv/f=-\vec{q}_C$.
Thus, to see the quadrature corresponding to the real part we should
choose
\begin{eqnarray}
  \label{eq:phiLO2ff-constant-real-not-optimized}
  \psi_{2,f}^\mathrm{LO}=
  \phi_{G}(-\vec{q}_C,0)-\phi_{1}^\mathrm{LO}(\vec{x}_1)+\pi.  
\end{eqnarray}
The quadrature component corresponding to the imaginary part of the
object far field object may consequently be observed by adding $\pi/2$
to this value.

We can improve this result by using that, as shown in
App.~\ref{sec:Append-appr-form}, the gain phase can be
approximated with a quadratic expansion in $\vec{q}$, i.e.
$\phi_{G}(\vec{q},0)\simeq
\phi_{G}^{(0)}+\vec{\phi}_{G,q}^{(1)}\cdot
\vec{q}+\phi_{G,q}^{(2)}|\vec{q}|^2 $ [see Eq.~(\ref{eq:phi_app})].
The linear term $\vec{\phi}_{G,q}^{(1)}$ can be compensated by making the
idler LO a tilted wave instead of a plane wave, i.e.
\begin{equation}
  \label{eq:phiLOf}
\phi_{2,f}^\mathrm{LO}(\vec{x}_2)=\psi_{2,f}^\mathrm{LO}
+\vec{q}_{2,f}^\mathrm{LO}\cdot\vec{x}_2 ,  
\end{equation}
where $\psi_{2,f}^\mathrm{LO}$ is a controllable reference phase and
the wave number $\vec{q}_{2,f}^\mathrm{LO}$ may be applied to the LO
by using a grating. We should therefore choose the following value of
the LO wave number
\begin{equation}
  \label{eq:qLOf}
  \vec{q}_{2,f}^\mathrm{LO}=-\vec{\phi}_G^{(1)}\kv/f=
\rho_2l_c\Psi_g\kv/f \vec{e}_x .
\end{equation}
The quadratic term $\phi_{G,q}^{(2)}$ may be cancelled by shifting the
focusing plane of the idler f-f system $\Delta z$ away from the crystal
plane $P$ \cite{brambilla:2004}, since this gives (in Fourier space)
a Fresnel transformation contribution of $\exp(-i\Delta
z|\vec{q}|^2/2\kv)$ on the gain function in Eq.~(\ref{eq:a1a2-q}).
Thus, using such a shift, the phase is approximately
\begin{eqnarray*}
  \phi_{G,\Delta z}(\vec{q},0)&\simeq&
  \phi_{G}^{(0)}+\vec{\phi}_{G}^{(1)}\cdot \vec{q}+(\phi_{G}^{(2)}-\Delta
  z/2\kv)|\vec{q}|^2 ,
\end{eqnarray*}
and by setting $\Delta z/2\kv= \phi_{G}^{(2)}$ the quadratic phase
term is exactly cancelled. So we must image a plane inside the crystal
since this choice from Eq.~(\ref{eq:phiG2}) implies
\begin{eqnarray}
  \label{eq:Deltaz}
  \Delta z=-(1/n_1+1/n_2)\Psi_gl_c,
\end{eqnarray}
with $\Psi_g$ defined in Eq.~(\ref{eq:psi_g}). Now all we need to do
is to find the overall reference phase. To observe the quadrature
component corresponding to the real part of the object far field we
should choose
\begin{eqnarray}
  \label{eq:phiLO2ff-constant-real}
\psi_{2,f}^\mathrm{
  LO}=\phi_G^{(0)}-\phi_{1}^\mathrm{LO}(\vec{x}_1)+\pi.
\end{eqnarray}
with $\phi_G^{(0)}$ given by Eq.~(\ref{eq:phiG0}). This result, the
tilted wave LO~(\ref{eq:qLOf}) and the shift of the focusing
plane~(\ref{eq:Deltaz}) represent the optimized scheme for accessing
the far-field image.

\subsubsection{Retrieval of object near field}
\label{sec:Retrieval-object-near}

The object near field can be observed by using the telescope setup in
the reference arm. The idler kernel is then found from (\ref{eq:h2_T})
and~(\ref{eq:hxq}) as
\begin{equation}
  \label{eq:Gamma2-2f}
  h_{2,\mathrm T}(\vec{x}_2,\vec{q})=(2\pi)^{-1}e^{-i\vec{q}
  \cdot \vec {x}_2}.
\end{equation}
Using this in~(\ref{eq:p12-mono}) we get
\begin{eqnarray}
p_T(\vec{x}_1,\vec{x}_2)= \frac{T_d
  \alpha_1^*(\vec{x}_1)\alpha_2^*(\vec{x}_2)}{2\pi i\lambda f}
  e^{-i\vec{x}_1\cdot\vec{x}_2\kv/f} 
\nonumber\\ 
\times
 \int
  \frac{d\vec{q}}{2\pi}e^{-i\vec{q}\cdot\vec{x}_2}  
\tilde T_\mathrm{
  obj} (\vec{q})G(\vec{x}_1\kv/f-\vec{q},0)+\mathrm{c.c.}
   \label{eq:p_T-semiexact}    
\end{eqnarray}
The correlation is therefore an integral over the gain function and
the object far field. Since the gain modulus is a roughly
plateau-shaped function in a region determined by $q_0$ that acts as a
cut-off of the higher values of $\vec{q}$, and the gain phase
dependence is slow within this region, the gain can as a good
approximation be pulled out of the integration as a constant evaluated
at the center of maximum gain $\vec{q}=-\vec{q}_C$. The remaining
integral is merely the inverse Fourier transform of the object
far field, i.e. simply the object near field and we have
\begin{subequations}
\begin{eqnarray}
  p_T(\vec{x}_1,\vec{x}_2)\simeq \frac{T_d|G(-\vec{q}_C,0)
  \alpha_1(\vec{x}_1)\alpha_2(\vec{x}_2)|}{\pi\lambda f}
\nonumber\\
\times \re[T_\mathrm{obj} (\vec{x}_2)e^{i\Phi_T(\vec{x}_1,\vec{x}_2)}]
 ,
  \label{eq:p_T}
\\
\Phi_T(\vec{x}_1,\vec{x}_2)\equiv -\vec{x}_2\cdot\vec{x}_1
\kv/f+\phi_G(-\vec{q}_C,0)
\nonumber\\
  -\phi_1^\mathrm{LO}(\vec{x}_1) -\phi_{2,T}^\mathrm{LO}(\vec{x}_2)
  - \pi/2 .
\label{eq:phi_T}
\end{eqnarray}
\end{subequations}
This is the main result for the retrieval of the object near field. We
should stress that it is only an approximate result that neglected a
cut-off of the high-frequency components of the object Fourier
transform. Thus, image information that relies on large
spatial frequency components will not be reproduced by the
correlations. 

If we can make $\Phi_T(\vec{x}_1,\vec{x}_2)$ constant and equal zero
or $-\pi/2$, the real or imaginary part of the near-field object is
observed. The linear contribution in $\vec{x}_2$ can be cancelled by
making the idler LO a tilted wave in the spirit of
Eq.~(\ref{eq:phiLOf}), i.e. $\phi_{2,T}^\mathrm{
  LO}(\vec{x}_2)=\psi_{2,T}^\mathrm{LO} +\vec{q}_{2,T}^\mathrm{
  LO}\cdot\vec{x}_2$ and choosing
\begin{eqnarray}
  \label{eq:qLO2f}
\vec{q}_{2,T}^\mathrm{ LO}=-\vec{x}_1\kv/f.  
\end{eqnarray}
In order to observe the real part of the object near field the overall
constant reference phase must be chosen to
\begin{eqnarray}
  \label{eq:phiLO22f-constant-real-no-opt}
  \psi_{2,T}^\mathrm{LO}=
  \phi_G(-\vec{q}_C,0) -\phi_{1}^\mathrm{LO}(\vec{x}_1)-\pi/2.
\end{eqnarray}

We may optimize the result~(\ref{eq:p_T}) by compensating the
quadratic term of the gain phase by taking the imaging plane inside
the crystal by the amount~(\ref{eq:Deltaz}), in the same way as
discussed in the previous section. In this case the idler reference
phase should instead be
\begin{eqnarray}
  \label{eq:phiLO22f-constant-real}
  \psi_{2,T}^\mathrm{LO}=
  \phi_G^{(0)} -\phi_{1}^\mathrm{LO}(\vec{x}_1)+ \vec{x}_1\cdot
  \vec{\phi}_G^{(1)}\kv/f-\pi/2. 
\end{eqnarray}
with $\phi_G^{(0)}$ and $\vec{\phi}_G^{(1)}$ given in
Eq.~(\ref{eq:gain-expansion}). 

This concludes the analytical part when point-like detectors are used
for measuring the signal quadrature components. We emphasize that the
retrieval of information about both the object near field and object
far field can be done by solely playing with the optical setup of the
reference arm. In the following subsection we show specific examples
from numerical simulations that support these results.

\subsection{Numerical results}
\label{sec:Numerical-results}

\subsubsection{General introduction}
\label{sec:General-introduction}

The numerical simulations presented in this paper take into account
the Gaussian profile as well as the pulsed character of the pump. They
consisted of solving the equations~(\ref{eq:waveq}) using the Wigner
representation to write the stochastic equations (see
\cite{brambilla:2004} for more details). At each pump shot a
stochastic field was generated at the crystal input with Gaussian
statistics corresponding to the vacuum state. Quantum expectation
values were obtained by averaging over pump shots. This simulates the
quantum behavior of symmetrically ordered operators, and thus the
correlations calculated are symmetrically ordered. The system we have
chosen to investigate is the setup of a current experiment performed
in Como \cite{jiang:2003,jedrkiewicz:2003}. Thus, we consider a BBO
crystal cut for type II phase matching. The pump pulse is at
$\lambda_0=352$ nm and has $\tau_0=1.5$ ps. The crystal length is
$l_c=4$~mm and the dimensionless gain parameter $\sigma_p l_c=4.0$.
Together with a pump waist of $w_0=660~\mu$m it corresponds roughly to
a pump pulse energy of 200 $\mu$J. The crystal is collinearly phase
matched at an angle $\theta=49.05^\circ$ of the pump propagation
direction with respect to the optical $z$ axis \cite{dmitriev:1999}.
We used $\theta=48.2^\circ$ which gives $\Delta_0<0$ and a gain that
is plateau shaped over a large region centered on $-\vec{q}_C$ as
given by Eq.~(\ref{eq:qC}).

The correlations typically converged after a couple of thousands
repeated pump shots when no spatial information was recorded in the
test beam (i.e. using either bucket or point-like detectors there).
For this reason it is very time-consuming to do a full 3+1D simulation
($xyt$ propagated along $z$). Instead we chose to model a 2+1D system:
a single transverse dimension $x$ along the walk-off direction and the
temporal dimension $t$ with propagation along the crystal direction
$z$. We will also show simulations where in addition to averaging over
shots, a spatial average over $\vec{x}_1$ is also performed. In the
2+1D simulations this reduces the number of shots needed by an order
of magnitude, while in 3+1D it is possible to average over both
transverse directions and a diffraction image can be observed even
with only a few shots.  Thus this method allows us to calculate the
correlations even from a full 3+1D simulation.

The integration along the crystal direction was done in $N_z=200$
steps. In the 2+1D simulations the $xt$-grid was chosen to $N_x=512$
and $N_t=32$, while for the 3+1D simulations we chose $N_x=N_y=256$
and $N_t=32$.  $(\vec{x},t)$ were scaled to $x_\mathrm{coh}$ and
$\tau_\mathrm{coh}$, respectively, as given by
Eq.~(\ref{eq:coherence-space-time}), while $z$ was scaled to the
crystal length $l_c$. 

We chose to investigate the quantitative behavior of the imaging system
with a transmitting double slit as the object, defined in one
transverse dimension as 
\begin{eqnarray}
  \label{eq:Tobj}
  T_\mathrm{obj}(x)=
\left\{  \begin{array}{ll}
0, & \;|x-\delta_x|>\frac{a+d}{2}, \; |x-\delta_x|<\frac{d-a}{2}\\
1, & \;\frac{d-a}{2}\leq |x-\delta_x|\leq \frac{a+d}{2}
  \end{array}
\right.,
\end{eqnarray}
where $d$ is the distance between the slit centers, $a$ is the slit
width, and $\delta_x$ is the object shift from origin. This object
only introduces amplitude modulations and does not alter phase
information. We take $\delta_x\neq 0$ in the numerics to compensate
for walk-off effects, so we must choose it properly in order for each
slit to see the same gain. The
analytical Fourier transform is
\begin{eqnarray}
  \label{eq:Pobj}
  \tilde T_\mathrm{obj}(q)&=&
a\sqrt{2/\pi}
e^{-i\delta_xq}\cos(qd/2)\mathrm{sinc}(qa/2),
\end{eqnarray}
where $\mathrm{sinc}(x)=\sin(x)/x$. The shift $\delta_x$ alone
introduces a nonzero imaginary part. Additionally we will present
results using a pure phase double slit \cite{abouraddy:2003}
\begin{eqnarray}
  \label{eq:Tobj-phase}
  T_\mathrm{obj}(x)=
\left\{  \begin{array}{ll}
-1, & \;|x-\delta_x|>\frac{a+d}{2}, \; |x-\delta_x|<\frac{d-a}{2}\\
1, & \;\frac{d-a}{2}\leq |x-\delta_x|\leq \frac{a+d}{2}
  \end{array}
\right..
\end{eqnarray}
This object does not alter amplitude information.

In homodyne detection the LO is typically taken from the same source
that created the pump. One would first generate the pump through
second-harmonic generation (SHG), and before the laser light at the
fundamental frequency enters the SHG crystal some of the power is
taken out using a beam splitter, and this field is then used for the
LOs. The second harmonic at the exit of the SHG process is then the
pump for the PDC process.  Therefore the LOs have more or less the
same energy, shape and duration as the pump pulse entering the PDC
setup.  Through numerical simulations with Gaussian LOs we confirmed
the results indicated in the analytical sections, namely that the
effect of a Gaussian shaped LO appears trivially as a
multiplication [see Eq.~(\ref{eq:p_f}) and~(\ref{eq:p_T})]. Since
broader LOs can easily be achieved experimentally using a beam
expander, we decided to keep the LOs as plane waves in space. Also
this allows to base the interpretation of the results on the physics
behind the PDC and keeping in mind that the LOs will only change the
result with a multiplication of a Gaussian function.  The temporal
duration was chosen identical to the pump (1.5 ps), which ensures a
good overlap between the pulses \cite{grosshans:2001}. 

\subsubsection{Averaging over shots}
\label{sec:Averaging-over-shots}

In this section we present numerical results from averaging over
repeated shots of the pump pulse, and where the optimization steps
discussed in the analytical section were carried out (i.e. cancelling
the quadratic and linear phase terms of the gain). We also present
semi-analytical calculations performed in Mathematica, which take the
analytical formulas from Sec.~\ref{sec:Analytical-results} and carry
out the optimization steps and in addition take explicitly into
account the finite gain.

The fixed position of the signal detectors $\vec{x}_1$ must be chosen
with care. In the far-field case we must ensure that in Eq.~(\ref{eq:p_f})
the object diffraction pattern $\tilde T_\mathrm{obj}
[(\vec{x}_1+\vec{x}_2)\kv/f]$ is centered where the gain modulus
$|G(-\vec{x}_2\kv/f,0)|$ has its maximum (which it has at
$-\vec{x}_2\kv/f=-\vec{q}_C$). Thus we must choose $x_1 \kv/f=-q_C=
-8.69$. As for the near-field setup Eq.~(\ref{eq:p_T-semiexact})
dictates that also here this position should be used to center the
gain over the origin of $\tilde T_\mathrm{obj}$. Thus, the test arm
detector position is unchanged as we pass from the far field to the
near field.

\begin{figure}[t]
\begin{center}
{\scalebox{.9}{\includegraphics{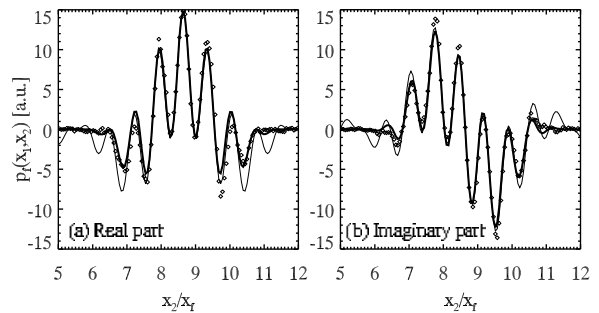}}}
\caption{The reconstruction of the object far-field distribution using
  the f-f setup in the idler arm.  The numerical signal-idler
  correlations (open diamonds) were calculated from averaging over
  2,000 repeated pump shots. In (a) and (b) $\psi_2^\mathrm{LO}$ was
  chosen as to observe the real and imaginary parts, respectively.
  The thin line shows $\tilde{T}_\mathrm{obj}$ from
  Eq.~(\ref{eq:Pobj}), while the thick line is Eq.~(\ref{eq:p_f}) as
  calculated in Mathematica. Slit parameters: $a=9$ pixels, $d=33$
  pixels and $\delta_x=23$ pixels, corresponding to $34~\mu$m,
  $123~\mu$m and $86~\mu$m, respectively, for $f=5~$cm.
  $x_f=fq_0/\kv$.}
\label{fig:farfield1}
\end{center}
\end{figure}

Let us first take the case of an f-f setup in the idler arm.
Figure~\ref{fig:farfield1} displays the two quadrature components
corresponding to the real and imaginary parts, and this confirms that
with this setup we can reconstruct the full phase and amplitude
information of the object diffraction pattern from the signal-idler
correlations. The open diamonds in the figure are results from the
numerical simulation. The results are compared to the analytical
result~(\ref{eq:Pobj}) shown with a thin line, as well as a
Mathematica calculation (thick line) that effectively plots
Eq.~(\ref{eq:p_f}). Compared to the analytical object distribution,
the numerical simulation shows that only the central part of the image
is reconstructed since the high frequency components die out. This is
a consequence of the finite gain bandwidth, as it was predicted by
Eq.~(\ref{eq:p_f}): the Mathematica calculation is based on this
equation and it almost coincides with the numerical results.  An
additional conclusion can be made from this agreement: since the
Mathematica result assumes that the pump is a stationary plane wave
this implies that the Gaussian profile in space and time of the pump
pulse in the numerics does not affect the correlations significantly.
This is because the object is in this example very localized so the
signal field impinging on it is almost constant.  At the end of the
section we shall see an example, the pure phase object, where this is
not the case.

\begin{figure}[t]
\begin{center}
{\scalebox{.9}{\includegraphics*{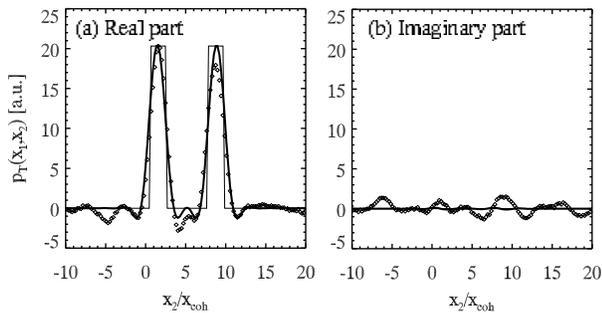}}}
\caption{The reconstruction of the object near-field distribution using
  the telescope setup in the idler arm. The notation and parameters
  are the same as in Fig.~\ref{fig:farfield1}, except the thin line
  here shows $T_\mathrm{obj}$ from Eq.~(\ref{eq:Tobj}) while the
  thick line is Eq.~(\ref{eq:p_T-semiexact}) as calculated in
  Mathematica. $x_\mathrm{coh}=1/q_0$.}
\label{fig:nearfield1}
\end{center}
\end{figure}

In Fig.~\ref{fig:nearfield1} we show that by merely exchanging the f-f
setup in the reference arm to the telescope setup, we can reconstruct
the object near-field distribution from the signal-idler correlations.
Fig.~\ref{fig:nearfield1} (a) shows the real part which follows the
shape of the analytical double slit~(\ref{eq:Tobj}), while the almost
zero imaginary part shown in (b) confirms that the object does not
alter phase information. The analytical SPWPA result using Mathematica
is calculated from Eq.~(\ref{eq:p_T-semiexact}) [note that it is the
integral over $\vec{q}$ that is calculated analytically and not the
approximated version given by Eq.~(\ref{eq:p_T})]. In comparison with
the analytical object (thin line) only the basic shape can be
recognized from the numerics, while as before the numerics agree very
well with the analytical Mathematica result. This again indicates that
the finite spatial bandwidth of the gain plays a decisive role in
reconstructing the object. Actually, the results from the near-field
and the far-field scenarios are linked: the far-field showed a cut-off
at high frequency components due to finite gain. Exactly this cut-off
in Fourier space is what gives the smooth character of the near-field
image: as predicted by Eq.~(\ref{eq:p_T-semiexact}) the correlation is
an integral over $\vec{q}$ of the gain multiplied with the object
far-field, and because the finite gain bandwidth acts as a low-pass
filter then it is not possible to reconstruct the sharp edges of the
double slit as this would require high-frequency spatial Fourier
components.

The ghost-imaging schemes distinguish themselves from the
Hanbury-Brown--Twiss (HBT) type schemes \cite{hanburybrown:1956} in
being able to pertain phase information of the image. In the HBT
technique the object is placed in both of the correlated beams. When
operated with thermal light the HBT technique can give information
about the Fourier transform of $|T_\mathrm{obj}|^2$, while when
operated with PDC beams it can give information about the Fourier
transform of $T_\mathrm{obj}^2$ \cite{abouraddy:2002,gatti:2004}.
Thus, the phase information is either lost or altered in the HBT
scheme. Instead, in the ghost-imaging schemes the object is only
placed in one of the beams, and from this one may instead reconstruct
the Fourier transform of $T_\mathrm{obj}$. This holds both for PDC
beams as well as for thermal (or thermal-like) beams
\cite{gatti:2004}.

\begin{figure}[t]
\begin{center}
{\scalebox{.9}{\includegraphics*{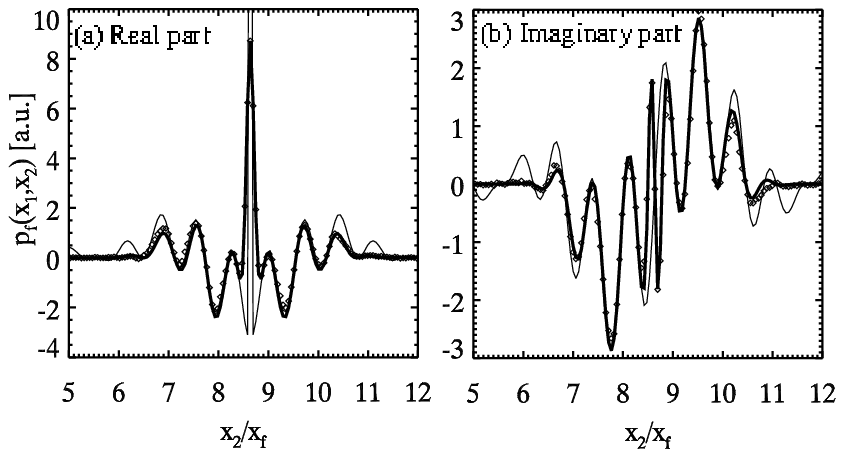}}}
{\scalebox{.9}{\includegraphics*{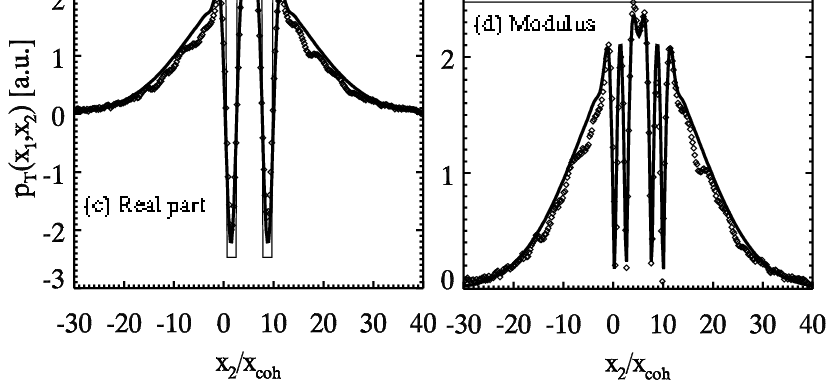}}}
\caption{The reconstruction of a pure phase double slit in (a)-(b) the f-f
  setup case, and (c)-(d) the telescope case. The notation and
  parameters are the same as in Figs.~\ref{fig:farfield1}
  and~\ref{fig:nearfield1}, except $10^4$ shots are used. Note that
  the Mathematica results (thick) have been adjusted to take into
  account the finite shape of the signal field impinging on the object.}
\label{fig:phase}
\end{center}
\end{figure}

As a specific demonstration we used the pure phase double
slit~(\ref{eq:Tobj-phase}) as an object, since for this object the HBT
scheme will not be able to provide information about the diffraction
pattern.  Figs.~\ref{fig:phase} (a) and (b) show the far-field
reconstruction with the f-f system in the idler arm, and both the real
and imaginary parts follow closely the analytical Fourier transform
(thin line). Thus, the scheme is able to reconstruct the Fourier
transform of a pure phase object. However, there is a disagreement for
the central peak in (a), corresponding to the average value of the
real part of the near field: the analytical value actually goes out of
the plot range shown.  This is due to the fact that ideally the
Fourier transform of the chosen phase object has a $\delta$-like
behavior here because all the photons are transmitted, whereas in the
numerics the signal field impinging on the object has a finite
extension. We accommodated for this in the Mathematica calculation by
multiplying the phase double slit with a Gaussian fitted to the
near-field profile of the signal in the numerics, and as can be seen
(thick line) the agreement is now very good.  In Figs.~\ref{fig:phase}
(c) and (d) the near-field reconstruction is shown with the telescope
setup in the idler arm, and as evidenced in the real part (c) the
object image can be successfully reconstructed (the imaginary part is
not shown since it was roughly zero). Also here the Mathematica
calculations include the same finite size of the signal field and
match very well with the numerics.  We should stress that the
``blurred'' shape of the slits is due to the finite gain bandwidth (as
in Fig.~\ref{fig:nearfield1}), while the deviation away from the slits
from the analytical result [thin line, Eq.~(\ref{eq:Tobj-phase})] is
due to the Gaussian profile of the signal [exactly this latter
behavior gives rise to a lower central peak in (a)]. Finally, we also
show the modulus of the reconstructed near field (d), and we do this
to stress that it actually reveals some features of the double slit.
This is a consequence of the finite gain bandwidth of the source which
turns the signal-idler fields from being completely incoherent (this
is the ideal result in the case of infinite bandwidth) to being
partially coherent. Note that in the numerics we had to use more shots
than before, otherwise the behavior away from the slits in the
near-field case was not too regular. This is a consequence of the more
noisy statistics there, as these parts have much lower intensities.

\subsubsection{The spatial average technique}
\label{sec:Averaging-over-shots-1}

In this section we describe how to improve the speed of the
correlation convergence as well as to obtain a much larger imaging
bandwidth.  The idea is to average the correlations not only over
repeated shots but also over the position of the signal detector in a
certain way. This implies either using a scanning point-like detector
setup or an array of detectors in the test arm. This abandons the idea
about keeping the test arm setup as simple as possible, but the
benefits of having an increased bandwidth and convergence rate should
not be neglected. The technique works for reconstructing the far
field, while for the near-field reconstruction this technique cannot
be applied for several reasons.  Most importantly, the final form of
the idler correlation~(\ref{eq:p_T}) when averaged over $\vec{x}_1$
gives the same result as using a bucket detector, and we already
argued in Sec.~\ref{sec:system-setup} through the unfolded Klyshko
picture that such a setup does not give any information of the object.
Another factor is that $\Phi_T(\vec{x}_1,\vec{x}_2)$ depends linearly
on $\vec{x}_1$, see Eq.~(\ref{eq:phi_T}) and~(\ref{eq:qLO2f}). Thus,
it is impossible to engineer the phases to observe a given quadrature
as $\vec{x}_1$ is varied.

For the f-f case, however, the technique is very powerful and it works
as follows. Let us take the case where we want to see the real part,
and assume we have performed all the optimization steps described in
Sec.~\ref{sec:Retrieval-object-far}.
The correlation is then
\begin{eqnarray}
  \label{eq:pfx1}
    p_f(\vec{x}_1,\vec{x}_2)\propto
\nonumber
|G(-\vec{x}_2\kv/f,0)\alpha_1(\vec{x}_1)\alpha_2(\vec{x}_2)|
\nonumber\\
\times\mathrm{Re}\left(\tilde T_\mathrm{obj}
    [(\vec{x}_1+\vec{x}_2)\kv/f]\right).
\end{eqnarray}
A change of coordinate system $\vec{x}\equiv
\vec{x}_1+\vec{x}_2$ and averaging over $\vec{x}_1$ as
$p_{f,\vec{x}_1}(\vec{x})\equiv \int d\vec{x}_1
p_f(\vec{x}_1,\vec{x})$ gives
\begin{eqnarray}
    p_{f,\vec{x}_1}(\vec{x}) \propto   \nonumber
\mathrm{Re}\left[\tilde T_\mathrm{obj}
    (\vec{x}\kv/f)\right]
\\\times
\int d\vec{x}_1 |G[(-\vec{x}+\vec{x}_1)\kv/f,0]\alpha_1(\vec{x}_1)
    \alpha_2(\vec{x}-\vec{x}_1)| 
,
\nonumber
\\
=\nonumber
\mathrm{Re}\left[\tilde T_\mathrm{obj}
    (\vec{x}\kv/f)\right]
\\\times
\int d\vec{\xi} |G[\vec{\xi}\kv/f,0]\alpha_1(\vec{\xi}+\vec{x})
    \alpha_2(-\vec{\xi})| 
,
  \label{eq:pfx1avg-xi}
\end{eqnarray}
where in~(\ref{eq:pfx1avg-xi}) we have made a change of integration
variable $\vec{\xi}\equiv \vec{x}_1-\vec{x}$. Assuming that 
$|\alpha_1(\vec{\xi}+\vec{x})|$ is slowly varying with respect to
$\vec{\xi}$ in the gain region for any $\vec{x}$ inside the object
diffraction pattern, we obtain
\begin{eqnarray}
    p_{f,\vec{x}_1}(\vec{x}) \propto   \nonumber
\mathrm{Re}\left[\tilde T_\mathrm{obj}
    (\vec{x}\kv/f)\right]|\alpha_1(\vec{x})|
\\\times
\int d\vec{\xi} |G[\vec{\xi}\kv/f,0]\alpha_2(-\vec{\xi})| ,
  \label{eq:pfx1avg}
\end{eqnarray}
The final integral is just a constant and does not depend on
$\vec{x}$, and hence as $\vec{x}$ is scanned the object far field is
observed multiplied by the modulus of the signal LO. Thus, whereas
with $\vec{x}_1$ fixed the imaging system had a finite bandwidth, now
the bandwidth is effectively only limited by the shape of the signal
LO (which consequently must be taken as broader than the far-field
diffraction pattern). Therefore, if we assume that
$|\alpha_1(\vec{x})|$ has a wide enough profile \textit{there is no
  cut-off of the spatial Fourier frequency components}. The
possibility of getting rid of the gain cut-off is is relying on the
object only being located in one arm so the gain is a function of
$\vec{x}_2$ only while the object far field is a function of
$\vec{x}_1+\vec{x}_2$. In contrast, if the object is located in both
arms, as in the HBT scheme, this would not be possible. Notice that
this procedure in practice does not amount to merely integrating over
$\vec{x}_1$ (as when a bucket detector is used). Rather, one should
move to position $\vec{x}_1$ of the signal detector while moving
together the position $\vec{x}_2$ of the idler detector as to keep
$\vec{x}=\vec{x}_1+\vec{x}_2$ constant. This corresponds to a spatial
convolution between the signal and idler quadratures. To see that,
consider the correlation of the measured quadratures $Z_j$, and from
Eq.~(\ref{eq:p12}) using the substitution
$\vec{x}=\vec{x}_1+\vec{x}_2$ we have
\begin{eqnarray}
    p_{f,\vec{x}_1}(\vec{x}) &=&\int_{T_d} dt_1 \int_{T_d} dt_2 
\nonumber\\&\times&
\int d\vec{x}_1  \la Z_1(\vec{x}_1,t_1) 
  Z_2(\vec{x}-\vec{x}_1,t_2) \ra 
\end{eqnarray}
In the numerical simulations this convolution is rapidly calculated
using the fast Fourier transform method.

We remark that the bandwidth can also be improved without applying the
spatial average.  Namely, if instead $\vec{x}_2$ is kept fixed and
$\vec{x}_1$ is scanned, according to Eq.~(\ref{eq:p_f}) the gain
cut-off is not present, while the diffraction pattern still emerges
from the correlations. We performed numerical simulations that showed
that this technique in fact gives an unlimited bandwidth as when a
spatial average is performed over $\vec{x}_1$. However, since no
spatial average is performed the convergence rate is as slow as when
$\vec{x}_1$ is kept fixed. If an average is done over $\vec{x}_2$ we
obtain the same result as the spatial average over $\vec{x}_1$. For a
more complete discussion on extending the imaging bandwidth see
Ref.~\cite{bache:2004a}.

\begin{figure}[t]
\begin{center}
{\scalebox{.9}{\includegraphics*{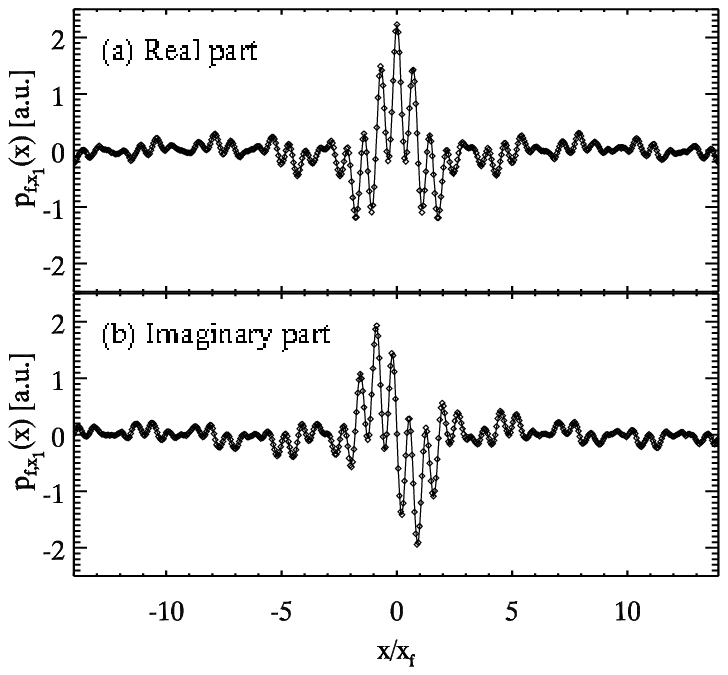}}}
{\scalebox{.9}{\includegraphics*{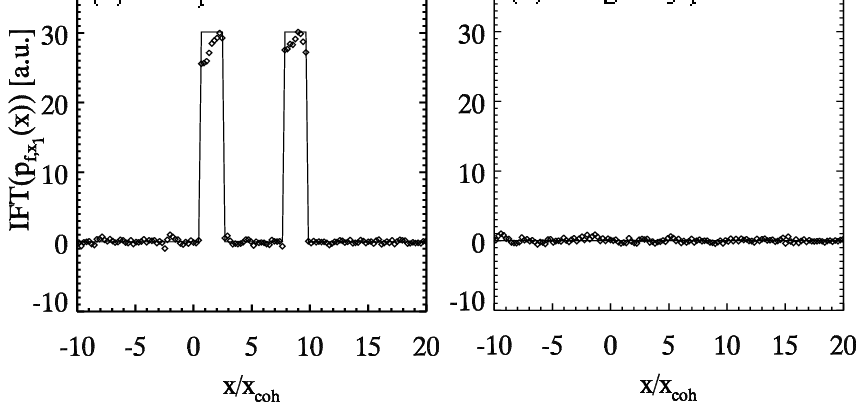}}}
\caption{The same as shown in Fig.~\ref{fig:farfield1} except the
  average is performed over 200 pump shots as well as over $x_1$. (a)
  and (b) show the real and imaginary parts of the reconstructed
  object far-field.  Notice that the full numerical grid is shown. (c)
  and (d) show the real and imaginary parts of the inverse Fourier
  transform of the reconstructed far-field correlation function. The
  thin lines display the analytical results of Eq.~(\ref{eq:Pobj}) for
  the far field and~(\ref{eq:Tobj}) for the near field.}
\label{fig:farfieldSM}
\end{center}
\end{figure}

With the same system setup as in Fig.~\ref{fig:farfield1}, the result
for averaging over all $x_1$ is shown in Fig.~\ref{fig:farfieldSM}.
First of all, we note the excellent agreement between the numerics and
the analytical far-field pattern, and also that there is no frequency
cut-off: the imaging bandwidth has become practically infinite.
Moreover, only 200 shots were needed to obtain the same degree of
convergence as in Fig.~\ref{fig:farfield1}, which is an order of
magnitude faster. Another interesting point about homodyne detection
is that by measuring both quadratures in the the far-field
distribution, we may reconstruct the complete near-field object
distribution from this information by using the inverse Fourier
transform. We have done this for the data in Fig.~\ref{fig:farfieldSM}
(a) and (b) and the result is shown in (c) and (d). The real part,
(c), of the inverse Fourier transform follows $T_\mathrm{obj}$ very
precisely.  This is because we now have access to many more high
frequency components compared to the case shown in
Fig.~\ref{fig:nearfield1}. Thus, as the far-field imaging bandwidth is
increased, the near-field resolution improves. It is instructive to
note that in the absence of the spatial average the inverse Fourier
transform of the far-field correlations shown in
Fig.~\ref{fig:farfield1} would give \textit{exactly} the result
reported with the telescope setup in Fig.~\ref{fig:nearfield1}. This
again underlines the strong link between the near-field and the
far-field measurements when all the phase information is intact.

\begin{figure}[t]
\begin{center}
{\scalebox{.9}{\includegraphics*{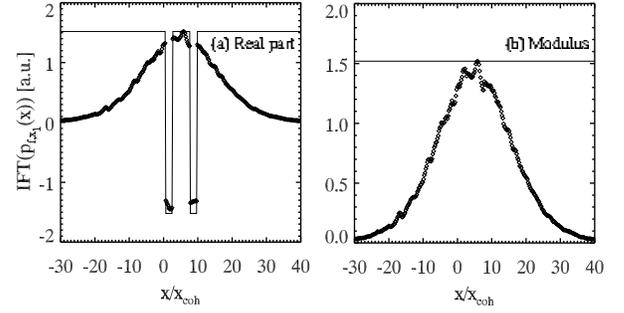}}}
\caption{The same setup as used in Fig.~\ref{fig:phase} except now the
  average is also done over $x_1$. The far-field quadrature
  correlations are calculated from 1,000 repeated pump shots, and (a)
  and (b) show the real part and modulus, respectively, of the
  inverse Fourier transform of the correlation.}
\label{fig:phaseSM}
\end{center}
\end{figure}

Returning to the phase object of Fig.~\ref{fig:phase} we saw that with
the signal detector position fixed the finite gain bandwidth makes the
imaging system partially coherent and some phase information is
present in the modulus of the correlations. For the phase double slit
we repeated the simulations using the spatial average technique to
reconstruct the far-field distribution. The inverse Fourier transform
of the correlation is shown in Fig.~\ref{fig:phaseSM} and the real
part now shows excellent agreement with the analytical phase double
slit (thin line). Most importantly, Fig.~\ref{fig:phaseSM}b showing
the modulus now does not reveal any information about the double slit.
The extended imaging bandwidth achieved with the spatial average makes
the imaging system truly incoherent and hence no phase information is
transferred to the modulus.

\begin{figure}[t]
\begin{center}
{\scalebox{1}{\includegraphics*{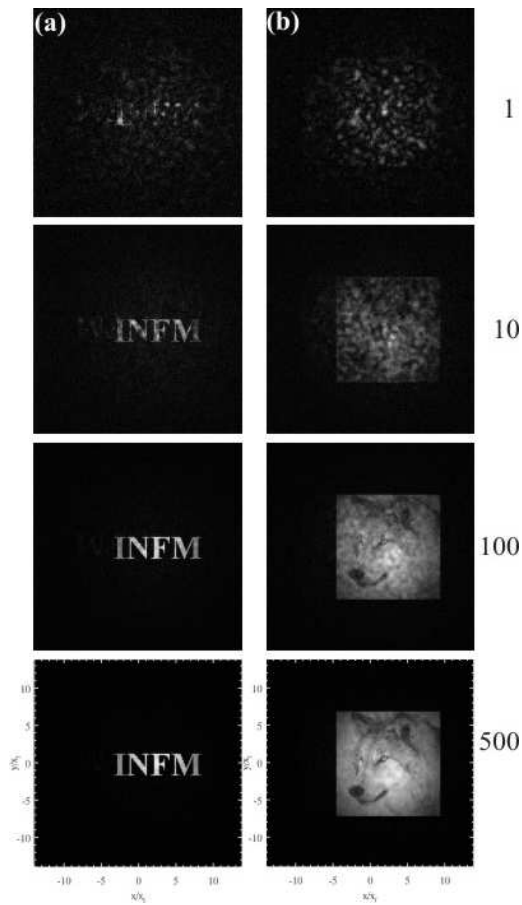}}}
\caption{The modulus of the inverse Fourier transform of the far-field
  correlation function as obtained in an f-f setup with two different
  objects (a) and (b). The correlations are calculated from a full 3+1D
  simulation, and averaging over the entire $\vec{x}_1$ plane and over
  the number of repeated pump shots shown on the right.}
\label{fig:farfieldSM2D}
\end{center}
\end{figure}

The spatial average over $\vec{x}_1$ works even better in two
transverse dimensions because there are many more points to average
over. In Fig.~\ref{fig:farfieldSM2D} we show two cases where different
objects were used: (a) a simple amplitude transmission mask with the
letters ``INFM'', and (b) a more complicated amplitude transmission
mask showing a picture of a wolf. The modulus of the inverse Fourier
transform of the far-field correlation is shown for different number
of shots. Evidently the simple mask (a) converges faster than the more
complicated mask (b), but nevertheless in both cases a good, sharp
image is obtained after relatively few shot repetitions.  After
additional averaging over shots the irregularities disappear, as
indicated in the last shots.

\begin{figure}[t]
\begin{center}
{\scalebox{0.7}{\includegraphics*{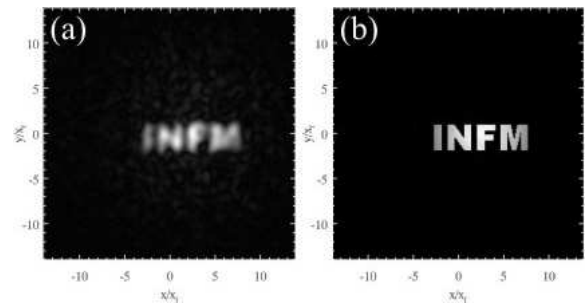}}}
\caption{Qualitative comparison of the resolution in (a) the fixed
  detector case and (b) the spatial average case. (a) uses the
  telescope setup in the idler arm to obtain the near-field
  correlation, and the modulus is shown. (b) uses the f-f setup in the
  idler arm to obtain the far-field correlation, and what is shown is
  the modulus of the inverse Fourier transform of the correlation. The
  simulations neglect the time dimension and 5,000 shots are used for
  the averages. }
\label{fig:2Dcompare}
\end{center}
\end{figure}

We now evaluate the qualitative difference of the near-field
resolution between the fixed detector case and the spatial average
technique. Fig.~\ref{fig:2Dcompare} (a) displays the modulus of the
near-field correlations obtained by using the telescope setup and a
fixed signal detector position, and Fig.~\ref{fig:2Dcompare} (b)
displays the modulus of the inverse Fourier transform of the far-field
correlations obtained using the f-f setup and applying the spatial
average technique. The blurred character in the fixed detector case
that was also present in one transverse dimension (see e.g.
Fig.~\ref{fig:nearfield1}) is now very obvious, and does not improve
much even if more averages are performed since it is a consequence of
the finite bandwidth as mentioned before.  In contrast the much
improved far-field bandwidth of the spatial average technique
increases the near-field resolution so the observed near-field image
completely sharp. Note that the simulations in
Fig.~\ref{fig:2Dcompare} have been performed neglecting the time
dimension, without which a dramatic extension in the amount of
computational time would occur. Such an approximation corresponds to
using a narrow interference filter, and the approach was justified by
comparing the correlations at a fewer number of shots with simulations
including also time.

We have in this section shown that using point-like detectors in the
test arm allows to reconstruct both amplitude and phase information
about an image. The near-field and far-field object distributions
could be obtained by only changing the optical setup of the reference
arm. The spatial imaging bandwidth for a fixed test detector position
is determined by the source bandwidth, and we showed that this in turn
determines the near-field resolution. Using a spatial average
technique improves the imaging bandwidth dramatically in the far-field
case, and also leads to a much faster information retrieval.

\section{Bucket detector setup in test arm}
\label{sec:Bucket-detector-at}

We now turn to the case where bucket detectors are used in the test
arm. One motivation is to show that by using homodyne detection
together with bucket detectors, phase-only objects can be imaged. If an
intensity detection scheme is used this is not possible, as shown in
Ref. \cite{bennink:2002a} for the two-photon coincidence imaging case
(although it is possible using a point-like detector in the object arm
\cite{abouraddy:2001,abouraddy:2003}). This is a general result that
also holds in the high-gain regime (it can easily be derived from the
results of Ref.\cite{gatti:2003,gatti:2004}).  Another motivation is
to see the imaging capabilities of this setup since it technically
seems simpler than the point-like detector setup.

\subsection{Analytical results}
\label{sec:Analytical-results-bucket}

As argued in Sec.~\ref{sec:system-setup}, when bucket detectors are
used in the test arm it is necessary to keep the object placed in the
measurement plane. Besides that the lens setups in the arms are
arbitrary, but we decided to keep the f-f setup in the signal arm, as
well as the idler setup in either f-f or the telescope configuration.
The test-arm set-up is shown in Fig.~\ref{fig:setup}c. The signal
kernel~(\ref{eq:h1-bucket-xx'}) is transformed according
to~(\ref{eq:hxq}) as
\begin{eqnarray}
  h_1^\mathrm{b}(\vec{x}_1,\vec{q})&=&\frac{2\pi}{i \lambda f}
  \delta(\vec{q}+\vec{x}_1\kv/f)T_\mathrm{obj}( \vec{x}_1).
  \label{eq:h1-bucket-xq}
\end{eqnarray}
The idler kernels remain unchanged. The bucket detectors in the test
arm effectively corresponds to an integration over $\vec{x}_1$ of the
signal-idler quadrature correlation as $\bar{ p}_T^\mathrm{
  b}(\vec{x}_2)\equiv \int d\vec{x}_1 p_T^\mathrm{
  b}(\vec{x}_1,\vec{x}_2)$.

\subsubsection{Retrieval of object near field}
\label{sec:Retrieval-object-far-bucket}

For an f-f setup in the reference arm the idler kernel is given by
Eq.~(\ref{eq:h2q-ff}) and from Eq.~(\ref{eq:p12-mono}) we therefore
obtain
\begin{eqnarray}
  p_f^\mathrm{
    b}(\vec{x}_1,\vec{x}_2)&=&\frac{T_d}{\pi}
    \delta(\vec{x}_1+\vec{x}_2)|
    \alpha_1(\vec{x}_1)\alpha_2(\vec{x}_2)G(\vec{x}_1\kv/f,0)  | 
\nonumber\\&\times& 
\re[T_\mathrm{obj} (\vec{x}_1) e^{i\Phi_f^\mathrm{
    b}(\vec{x}_1,\vec{x}_2)}], 
\end{eqnarray}
The measured signal-idler correlation is consequently
\begin{subequations}
\begin{eqnarray}
  \bar{ p}_f^\mathrm{
    b}(\vec{x}_2)&=&
\frac{T_d}{\pi}
|\alpha_1(-\vec{x}_2)\alpha_2(\vec{x}_2)G(-\vec{x}_2\kv/f,0)  | 
\nonumber\\&\times &
\re[T_\mathrm{obj} (-\vec{x}_2)
e^{\bar{\Phi}_f^\mathrm{b}(\vec{x}_2)}],
  \label{eq:p_f-bucket-int}
\\
\bar{\Phi}_f^\mathrm{b}(\vec{x}_2)&\equiv& \phi_G(-\vec{x}_2\kv/f,0) 
\nonumber\\
&&-\phi_1^\mathrm{LO}(-\vec{x}_2) -\phi_{2,f}^\mathrm{
  LO}(\vec{x}_2) +\pi  .
\label{eq:phi_f-bucket}
\end{eqnarray}
\end{subequations}
This is the main result for retrieving the near field in the bucket
detector case. 

The idler LO phase may then be engineered to observe the desired
quadrature in much the same way as shown in in
Sec.~\ref{sec:Retrieval-object-far}. As a first approximation the gain
phase dependence is neglected, so to observe the real part we must have
$\bar{\Phi}_f^\mathrm{b}(\vec{x}_2)=0$, implying we should choose
\begin{eqnarray}
  \label{eq:phiLO2ff-constant-real-not-optimized-bucket}
  \psi_{2,f}^\mathrm{LO}=
  \phi_{G}(-\vec{q}_C,0)-\phi_{1}^\mathrm{LO}(-\vec{x}_2)+\pi.  
\end{eqnarray}

We may optimize this result by setting the focus plane of the f-f
setup in the idler arm inside the crystal with the amount given by
Eq.~(\ref{eq:Deltaz}) as to compensate the quadratic term of the gain
phase. Additionally, by making the idler LO a tilted wave with the
wave number given by Eq.~(\ref{eq:qLOf}) we compensate for the linear
phase term. With these optimizations, the real part can be observed by
choosing
\begin{eqnarray}
  \label{eq:phiLO2ff-constant-real-bucket}
  \psi_{2,f}^\mathrm{LO}=\phi_G^{(0)}
-\phi_{1}^\mathrm{LO}(-\vec{x}_2)+\pi.  
\end{eqnarray}

We note that the results show a dependence of the signal LO as
$\alpha_1(-\vec{x}_2)$. Therefore, as the idler pixels are scanned it
is crucial that the signal LO does not vary substantially. Another
point is the scale of the coordinate system, but we will return to
this in the next subsection.

\subsubsection{Retrieval of object far field}
\label{sec:Retr-object-far-bucket}

For a telescope setup in the reference arm the idler kernel is given
by~(\ref{eq:Gamma2-2f}) and thus using Eq.~(\ref{eq:p12-mono}) we have
\begin{eqnarray}
  p_T^\mathrm{
    b}(\vec{x}_1,\vec{x}_2)=\frac{T_d}{2\pi i\lambda f}
\alpha_1^*(\vec{x}_1)\alpha_2^*(\vec{x}_2)G(\vec{x}_1\kv/f,0)
\nonumber\\\times 
T_\mathrm{obj} (\vec{x}_1)
e^{-i\vec{x}_1\cdot\vec{x}_2\kv/f}+\mathrm{c.c.}
\end{eqnarray}
Thus, the measured correlation is
\begin{eqnarray}
  \bar{ p}_T^\mathrm{ b}(\vec{x}_2)&=&
\frac{T_d}{i\lambda f}\alpha_2^*(\vec{x}_2)
\nonumber\\&\times&
\int \frac{d\vec{x}_1}{2\pi} e^{-i\vec{x}_1\cdot\vec{x}_2\kv/f}
    G(\vec{x}_1\kv/f,0)   
    \alpha_1^*(\vec{x}_1) T_\mathrm{obj} (\vec{x}_1)
\nonumber\\
&&+\mathrm{c.c.}
\label{eq:p_T-bucket-int}
\end{eqnarray}
We assume now that the plateau-shaped gain and the modulus of the
signal LO vary slowly with respect to the object, so they can be taken
out of the integral, evaluated at the position of maximum gain
$\vec{x}_1\kv/f=-\vec{q}_C$. This approximation yields
\begin{subequations}
\begin{eqnarray}
  \bar{ p}_T^\mathrm{
    b}(\vec{x}_2)\simeq 
\frac{2T_d}{\lambda f}|G(-\vec{q}_C,0) \alpha_1(-\vec{q}_Cf/\kv)
    \alpha_2(\vec{x}_2)| 
\nonumber\\\times 
\re[\tilde T_\mathrm{obj} (\vec{x}_2\kv/f+\vec{q}_1^\mathrm{LO})
    e^{i\bar{\Phi}_T^\mathrm{
    b}(\vec{x}_2)}], 
  \label{eq:p_T-bucket}
\\
\bar{\Phi}_T^\mathrm{b}(\vec{x}_2)\equiv 
\phi_G(-\vec{q}_C,0)
-\psi_1^\mathrm{LO} -\phi_{2,T}^\mathrm{
  LO}(\vec{x}_2) +\pi/2  .
\label{eq:phi_T-bucket}
\end{eqnarray}
\end{subequations}
This is the main result showing that the object far field can be
reconstructed in the bucket detector case. 

We can observe the real of the far-field of the object if
$\bar{\Phi}_T^\mathrm{b}(\vec{x}_2)=0$, which in the case
of~(\ref{eq:phi_T-bucket}) implies
\begin{equation}
  \label{eq:phi-bucket-T-not-optimized}
\phi_{2,T}^\mathrm{
  LO}(\vec{x}_2)=  \phi_G(-\vec{q}_C,0)
-\psi_1^\mathrm{LO} +\pi/2.
\end{equation}
As in the other cases the result~(\ref{eq:p_T-bucket}) can be
optimized by compensating for the quadratic gain phase term before the
integration by taking the imaging plane of the telescope setup inside
the crystal with the amount given by Eq.~(\ref{eq:Deltaz}). In that
case we should choose the following reference phase for the idler LO
to observe the real part
\begin{equation}
  \label{eq:phi-bucket-T-optimized}
\phi_{2,T}^\mathrm{
  LO}(\vec{x}_2)=  \phi_G^{(0)} 
-\psi_1^\mathrm{LO} +\pi/2.
\end{equation}
In connection with optimizing the imaging performance, it is apparent
that if the signal LO is a tilted wave then the origin of the
reconstructed diffraction pattern changes. This can be used to
compensate for undesired walk-off effects by centering the
reconstructed image in the place where the idler (which in this setup
is in the near field) has its maximum. Additionally, making the idler
LO a tilted wave can be used to compensate for unimportant
oscillations in the quadrature components that comes from shifting the
object near field from origin [see in that connection Eq.~(\ref{eq:Pobj})].

Let us come back to the scaling of the image information. The
particular setup we chose (having the object placed in the far field
created by an f-f lens system, just before the bucket detectors)
implies that the signal field impinging on the object changes on the
scale $x_f=f q_0/\kv$. This is evident in the analytical formulas
Eqs.~(\ref{eq:phi_f-bucket}) and~(\ref{eq:p_T-bucket-int}). In
contrast, when the object was placed in the near field the field
changed on the scale $x_\mathrm{coh}=1/q_0$. As an example, for
$f=5$~cm and $\lambda=704$~nm, $x_f=338~\mu$m while
in contrast $x_\mathrm{coh}=17~\mu$m. The consequence is that while
imaging is possible with bucket detectors, it occurs on a different
(i.e. larger) length scale than with point-like detectors.

\subsection{Numerical results}
\label{sec:Numerical-results-bucket}

The numerical simulations in the bucket detector case were done
exactly as discussed in Sec.~\ref{sec:Numerical-results}, except now
the object is placed after the f-f setup in the test arm. We decided
to keep the object structure used in the point-like detector case,
i.e. exactly the same number of pixels between the slits as well as
for the slit aperture. As discussed above, this means that the slit
dimensions are determined by the larger scale $x_f$.  Moreover, the
pure phase double slit in one transverse dimension is used as an
object in order to demonstrate the homodyne detection scheme's
capabilities. Finally, we mention that in the numerics all the
optimization procedures discussed in the previous sections were
carried out. This includes in the telescope case to use a tilted wave
signal LO to center the reconstructed diffraction pattern over the
peak of the idler near field, as well making the idler LO a tilted
wave as to cancel the oscillations in the diffraction pattern
originating from centering the object over the signal far-field
plateau. These latter two optimizations are not crucial for the
imaging performance.

\begin{figure}[t]
\begin{center}
{\scalebox{.9}{\includegraphics*{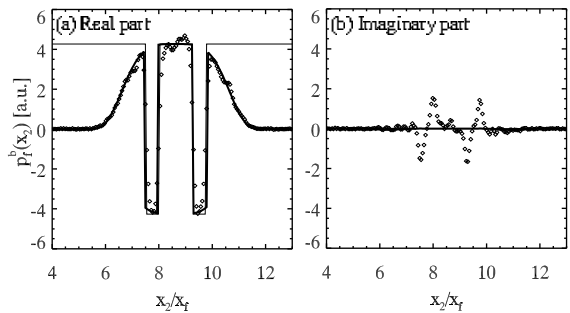}}}
\caption{The reconstruction of the object near-field distribution of a
  pure phase object with bucket detectors in the test arm and using
  the f-f setup in the reference arm. The numerical correlations (open
  diamonds) were calculated from averaging over $10^4$ pump shots. The
  thin line is the analytical phase double slit, while the thick line
  is calculated in Mathematica on basis of
  Eq.~(\ref{eq:p_f-bucket-int}). Slit parameters in pixels as
  Fig.~\ref{fig:farfield1} corresponding for $f=5$~cm to $a=166~\mu$m,
  $d=610~\mu$m and $\delta x=-2.9~$mm.
}
\label{fig:nearfield-bucket}
\end{center}
\end{figure}

Fig.~\ref{fig:nearfield-bucket} shows the numerical simulation of the
reconstruction of the object near field using the f-f setup in the
idler arm. The reconstructed real part in (a) follows the analytical
double slit, but notice that the Mathematica result (thick line) has a
better resolution than the numerical results. This indicates that the
finite shape of the pump worsens the resolution, and we will come back
to this later. The imaginary part (b) is containing
only little information about the slits. Away from the slits the
correlations decay to zero because of the finite shape of the
far-field gain.

\begin{figure}[t]
\begin{center}
{\scalebox{.9}{\includegraphics*{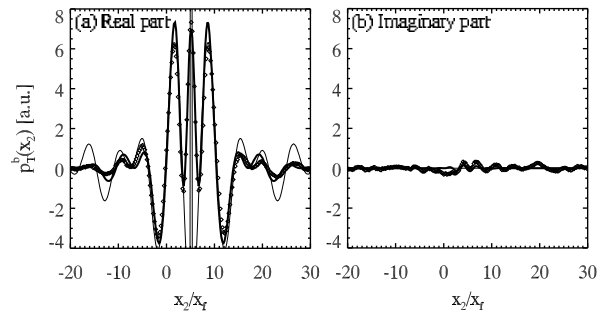}}}
\caption{The same as shown in Fig.~\ref{fig:nearfield-bucket} except
  now using the telescope setup in the idler arm making it possible to
  reconstruct the object far-field distribution. The thick line is
  Eq.~(\ref{eq:p_T-bucket-int}) calculated in Mathematica while taking
  into account the finite shape of the idler near-field distribution.
}
\label{fig:farfield-bucket}
\end{center}
\end{figure}

Fig.~\ref{fig:farfield-bucket} shows a numerical simulation similar to
Fig.~\ref{fig:nearfield-bucket} but with the telescope setup in the
idler arm allowing for the reconstruction of the object far-field
distribution. The reconstructed real part in (a) follows within a
certain bandwidth the analytical result (thin line). The cut-off is
actually caused by the Gaussian shape of the idler near-field profile,
which has been taken into account in the Mathematica calculation of
the integral~(\ref{eq:p_T-bucket-int}); without such a correction the
Mathematica result would actually very closely follow the analytical
result (thin line) without being cut off. This can be understood by
inspecting Eq.~(\ref{eq:p_T-bucket-int}) more closely, since we see
that the action of the gain inside the integral is to provide a limit
for the object extension. However, in contrast to the point-like
detector case of Eq.~(\ref{eq:p_f}) where the gain eventually
determines the far-field imaging bandwidth, in the bucket case it is
not so: Once the object is located inside the gain, the reproduced
diffraction pattern is very good. This is also in accord with what we
saw in the semi-analytical reproduction of the near field in
Fig.~\ref{fig:nearfield-bucket}, where the double slits appear very
sharp indicating that the Fourier frequency bandwidth is large. A
consequence of this is that if we tailor the pump to have a plane-wave
shape, we should see practically no cut-off in the far field as well
as a very sharp reproduction of the double slit in the near field. We
checked this to be the case. Thus, also the bucket detector case shows
an inherent link between far-field bandwidth and near-field
resolution. Finally, note that the imaginary part (b) is zero because
we chose to eliminate the oscillations arising because of the offset
of the object from origin, making the Fourier transform purely real.
Additionally, the central peak of the analytical function in (a) goes
out of the range shown for the same reasons as in
Fig.~\ref{fig:phase}.

In conclusion, a homodyne detection scheme with bucket detectors in
the test arm is capable of observing both amplitude and phase
distribution of an object. Both the near-field and the far-field
distributions are accessible by only changing the optical setup in the
reference arm and even phase-only objects can be imaged.

\section{Conclusion}
\label{sec:Conclusion}

We have shown how a homodyne detection technique can be used to get
access to both modulus and phase information in ghost-imaging schemes
based on parametric down-conversion in the high-gain regime. It is
necessary to measure both the signal and idler beams with individual
local oscillators, and we have shown analytically how to engineer the
phases of these in order to retrieve both the object image (near
field) and the object diffraction pattern (far field). The results
were confirmed with numerical simulations taking into account a finite
shape of the pump pulse.

We showed that the resolution in the near field and the imaging
bandwidth in the far field were inherently linked, and were restrained
to the finite gain of the source. Thus in the far-field setup the gain
cuts off the higher frequency components in the spatial Fourier
domain. In the near-field setup the image is more blurred than the
reference image, and we showed that this is exactly a consequence of
the higher frequency components missing in the spatial Fourier domain
because of the gain cut-off.

The homodyne scheme allows us to have access to both phase and
amplitude information, implying it is possible to pass from the
far-field result to the near-field result by simply making an inverse
Fourier transform. Thus by taking the far-field correlations
containing the information about the diffraction pattern and
performing an inverse Fourier transform gave the same result as
measuring the near-field correlations containing the information about
the object image, and vice versa. This also means that once we have
measured e.g. the far field there is actually no need to also measure
the near field.

We showed that in the far-field setup a spatial average technique over
the signal detector pixels leads to much faster convergence rate of
the correlations (fewer pump-shot repetitions needed) as well as a
hugely improved bandwidth of the diffraction pattern. An intuitive
explanation of the method is that each far-field mode is uncorrelated
to the rest of the modes in the beam, but the strong mode correlation
between the signal-idler beams means that for each signal mode an
independent correlation can be measured containing the object
information. The spatial average then corresponds to averaging over
all the modes in the signal beam.  This means that much fewer repeated
pump shots of the laser are needed before the correlations converge,
and this truly exploits the possibility for parallel operations in
spatially correlated beams. The increased bandwidth stands out as the
true advantage of the technique, however, and it relies on the fact
that as we change the test detector position a different part of
diffraction pattern is amplified. Thus, as we average over all the
positions we effectively also extract information about different
parts of the diffraction pattern, resulting in an increased bandwidth.
It is important to note that the technique also works when using
direct intensity measurements of the fields instead of homodyne
detection, both when using PDC beams (as will be discussed elsewhere
in a separate publication \cite{bache:2004a}) and the thermal-like
beams investigated in \cite{gatti:2004}.  The technique works
particularly well in the high-gain regime because the number of signal
photons per mode is not small, but should also work in the low-gain
regime.

In the time domain the outcome of a homodyne detection is depending
strongly on the temporal overlap between the local oscillator and the
field to be measured. A question that had to be answered in this paper
was therefore how important it is to ensure a proper overlap also in
the spatial domain between the local oscillator and the field. The
answer in the context of ghost imaging is that it is not crucial. In
fact, the analytical results in the plane-wave pump approximation
suggest that the near-field shape of the local oscillator is merely
multiplied onto the final result. Even the numerical simulations
taking a Gaussian shape of the fields into account suggest the same.
Another issue entirely is measurements where the quantum efficiency is
important (in contrast to here), and in that case undoubtably the
spatial overlap becomes more important.

Using an intensity detection scheme the detection time must not be
larger than the coherence time, otherwise the background term that
contains no image information becomes dominating and the visibility of
the image is dramatically decreased
\cite{saleh:2000,gatti:2004,gatti:2004b}. In contrast, the homodyne
detection scheme is not restricted by this because the correlation
function is second order in the field operators implying there is no
background term. Hence the detection time may be chosen much larger
than the coherence time. Additionally, we have shown that the spatial
average technique can give a much larger imaging bandwidth than the
source itself. This suggests the possibility of using an optical
parametric oscillator (OPO) as a source, and using homodyne detection
as a measurement protocol. The limited spatial bandwidth of the OPO
with respect to PDC could then be circumvented with the spatial
average technique, and the homodyne measurements allow a cw operation
of the imaging scheme: since there are no problems with image
visibility, the measurement time can be taken longer than the
coherence time. Thus, one could effectively abandon the involved
scheme of using a high-power laser pumping a PDC setup and the
complicated task of overlapping LO fields with the signal-idler fields
with temporal durations on a pico-second level.

\section{Acknowledgments}
\label{sec:Acknowledgments}

This project has been carried out in the framework of the FET project
QUANTIM of the EU, of the PRIN project MIUR ``Theoretical study of
novel devices based on quantum entanglement'', and of the INTAS
project ``Non-classical light in quantum imaging and continuous
variable quantum channels''. M.B. acknowledges financial support from
the Danish Technical Research Council (STVF).

\appendix

\section{The approximate form of the gain phase}
\label{sec:Append-appr-form}

It is useful to obtain an approximate expression for the gain phase
$\phi_G$ from the expressions~(\ref{eq:uv}). We note in this
connection that
$\Delta_{12}(\vec{q},\Omega)=\Delta_{21}(-\vec{q},-\Omega)\equiv
\Delta(\vec{q},\Omega)$ and therefore also
$\Gamma_{12}(\vec{q},\Omega)=\Gamma_{21}(-\vec{q},-\Omega)\equiv
\Gamma(\vec{q},\Omega)$. Thus, we obtain
\begin{eqnarray}
  \label{eq:phase}
  \phi_{G}(\vec{q},\Omega)&=&-\Delta_0l_c+
\arctan\left[\frac{
    \Delta(\vec{q},\Omega) 
\tanh[\Gamma(\vec{q,\Omega})l_c]}{2\Gamma(\vec{q},\Omega)}\right]
\label{eq:phase1}\nonumber
\\
&\simeq&-\Delta_0l_c+\arctan\left[ \Psi_g\Delta(\vec{q},\Omega)l_c\right]
\nonumber\label{eq:phase2}
\\
&\simeq&-\Delta_0l_c+
\Psi_g\Delta(\vec{q},\Omega)l_c.
\label{eq:phase3}
\end{eqnarray}
The approximation to Eq.~(\ref{eq:phase3}) is that the gain is large
and that the modes inside the gain are phase matched
($\Delta(\vec{q},\Omega)\simeq 0$) making
$\Gamma(\vec{q},\Omega)\simeq \sigma_p$. We have also introduced a
dimensionless phase parameter that is related to the dimensionless
gain parameter $\sigma_pl_c$
\begin{equation}
  \label{eq:psi_g}
  \Psi_g\equiv\frac{\tanh\left(\sigma_pl_c\right)}{2\sigma_p l_c}.
\end{equation}
For $\sigma_pl_c\rightarrow 0$ (small gain) $\Psi_g\rightarrow 1/2$
while for $\sigma_pl_c\gg 1$ (large gain) $\Psi_g\rightarrow 0$. Now
from Eq.~(\ref{eq:Deltaq}) it is evident that $\Delta(\vec{q},\Omega)$
is quadratic in $\vec{q}$ and $\Omega$, so that the gain
phase~(\ref{eq:phase3}) approximately can be written as
\begin{subequations}
\label{eq:gain-expansion}
\begin{eqnarray}
  \label{eq:phi_app}
  \phi_{G}(\vec{q},\Omega)&\simeq&
  \phi_{G}^{(0)}+\vec{\phi}_{G,q}^{(1)}\cdot
  \vec{q}+\phi_{G,q}^{(2)}|\vec{q}|^2 
\nonumber\\
  &+&\phi_{G,\Omega}^{(1)}\Omega +\phi_{G,\Omega}^{(2)}\Omega^2
,\\
\label{eq:phiG0}
\phi_{G}^{(0)}&=&\Delta_0
  l_c(-1+\Psi_g)  
,\\
\label{eq:phiG1}
\vec{\phi}_{G,q}^{(1)}&=&-\rho_2  l_c\Psi_g\vec{e}_x
,\\
\label{eq:phiG2}
\phi_{G,q}^{(2)}&=&-\frac{(n_1+n_2)l_c\Psi_g}{2n_1n_2\kv}   
,\\
\label{eq:phiG1omega}
\phi_{G,\Omega}^{(1)}&=&
(k_1'-k_2')l_c\Psi_g
,\\
\label{eq:phiG2omega}
\phi_{G,\Omega}^{(2)}&=&
(k_1''+k_2'')l_c\Psi_g/2,
\end{eqnarray}
\end{subequations}
where $\vec{e}_x$ is a unit vector in the $x$-direction. We shall use
these expressions when the optimal phases of the LOs come into
question.

\section{The signal-idler correlation}
\label{sec:sign-idler-corr}

Here we present the detailed calculations of the general expressions
for the signal-idler correlation~(\ref{eq:p12_c}).
  
By using Eq.~(\ref{eq:c1c2-u1u2}) we see that in the signal-idler
correlation~(\ref{eq:p12_c}) the evaluation of the temporal part only
concerns the gain and the LOs, while the spatial image information
comes out from the integral over $\vec{q}$ in
Eq.~(\ref{eq:c1c2-u1u2}). To evaluate the temporal part of the
correlation we rewrite (\ref{eq:p12_c}) as
\begin{subequations}
\begin{eqnarray}
  \label{eq:pG}
&  p(\vec{x}_1,\vec{x}_2)=\int
  d\Omega B(\vec{x}_1,\vec{x}_2,\Omega) 
\nonumber\\
&\times\int d\vec{q}
h_1(\vec{x}_1,-\vec{q}) h_2(\vec{x}_2,\vec{q})G(\vec{q},\Omega) 
  +\mathrm{c.c.},
\\
&B(\vec{x}_1,\vec{x}_2,\Omega)\equiv T_d^2\int\int\frac{d\Omega'
  d\Omega''}{(2\pi)^2} 
  \alpha_1^*(\vec{x}_1,\Omega')\alpha_2^*(\vec{x}_2,\Omega'')   
\nonumber\\
&\times
\mathrm{sinc}[(\Omega-\Omega')T_d/2]
\mathrm{sinc}[(\Omega+\Omega'')T_d/2].
\label{eq:B}
\end{eqnarray}
\end{subequations}

Let us consider the case of a \textit{pulsed} LO, i.e. evaluate
Eq.~(\ref{eq:B}) for the LO duration much smaller than $\tau_\mathrm{
  coh}$.  Additionally taking into account that, as mentioned before,
it can be assumed that $T_d\gg \tau_\mathrm{coh}$ we may use the
approximation $T_d\mathrm{sinc}[(\Omega\pm\Omega')T_d/2]\simeq 2\pi
\delta(\Omega\pm\Omega')$.  In this limit we therefore obtain
\begin{eqnarray}
  \label{eq:B-pulse}
  B(\vec{x}_1,\vec{x}_2,\Omega)=
  \alpha_1^*(\vec{x}_1,\Omega)\alpha_2^*(\vec{x}_2,\Omega)   ,
\end{eqnarray} 
which means that the correlation is
\begin{eqnarray}
  \label{eq:p12-pulsed}
  p(\vec{x}_1,\vec{x}_2)= \int d\Omega
\alpha_1^*(\vec{x}_1,\Omega)\alpha_2^*(\vec{x}_2,\Omega)
\nonumber\\\times
  \int d\vec{q} 
h_1(\vec{x}_1,-\vec{q}) h_2(\vec{x}_2,\vec{q})  G(\vec{q},\Omega) 
  +\mathrm{c.c.}
\end{eqnarray}

In the case of a \textit{continuous wave} LO, the LO duration is much
longer that $\tau_\mathrm{coh}$. Thus, the LO corresponds to a
quasi-monochromatic wave, i.e.
$\alpha_j(\vec{x},\Omega)=\delta(\Omega)\alpha_j(\vec{x})$. Using
this form in Eq.~(\ref{eq:B}) removes the integration on $\Omega'$ and
$\Omega''$.  Then we use that for $T_d\gg\tau_\mathrm{coh}$ the
remaining term $T_d[\mathrm{sinc}(\Omega T_d/2)]^2$ behaves like
$2\pi\delta(\Omega)$, implying
\begin{eqnarray}
  \label{eq:B-mono}
  B(\vec{x}_1,\vec{x}_2,\Omega)\simeq\delta(\Omega)
  \alpha_1^*(\vec{x}_1)\alpha_2^*(\vec{x}_2)  T_d/(2\pi),
\end{eqnarray} 
which means that the correlation is
\begin{eqnarray}
  \label{eq:p12-mono-app}
  p(\vec{x}_1,\vec{x}_2)\simeq\frac{T_d}{2\pi}
\alpha_1^*(\vec{x}_1)\alpha_2^*(\vec{x}_2)
\nonumber\\
\times
  \int d\vec{q} 
h_1(\vec{x}_1,-\vec{q}) h_2(\vec{x}_2,\vec{q})  G(\vec{q},0) 
  +\mathrm{c.c.}
\end{eqnarray}

It is clear that the pulsed case may have a larger gain than the
monochromatic case because of the integration over $\Omega$. However,
a crucial point to achieve this goal lies in the minimization of the
$\Omega$-dependence of the phase in the integrand of
Eq.~(\ref{eq:p12-pulsed}). This will in turn maximize the integral
regardless of the value of $\vec{q}$. This can be done approximately
by putting in a proper temporal delay between the LOs, effectively
cancelling the first-order term~(\ref{eq:phiG1omega}) of the gain
phase. The second-order term cannot easily be cancelled, but we
estimate that this term does not contribute much since we have a
negligible contribution of this term of
$(k_1''+k_2'')l_c\Omega_0^2/2=(k_1''+k_2'')/[(k_2'-k_1')^2l_c]=O(10^{-4})$
for the crystal setup we consider (see
Sec.~\ref{sec:General-introduction} for details on the setup).
Numerical simulations showed that a LO optimized in this way as
expected enlarged the effective spatial bandwidth. However, the
improvement is not substantial and furthermore it requires temporally
short pulses (on the order of 20\% of the coherence time, or less)
which are difficult to obtain experimentally.  We therefore decided
only to show analytical results in the limit of a continuous wave LO.

\bibliographystyle{d:/LocalTexMf/miktex/prsty}
\bibliography{d:/Projects/Bibtex/literature}

\end{document}